\documentclass[sigconf, screen]{acmart}

\setcopyright{acmlicensed}
\usepackage{balance}       % to better equalize the last page
\usepackage{graphics}      % for EPS, load graphicx instead 
\usepackage{color}
\usepackage{booktabs}
\usepackage{multirow}
\usepackage{caption}
\usepackage{subcaption}
\usepackage{bbm}
\usepackage{pifont}
\usepackage{tabularray}
\usepackage{colortbl}

\newcommand{\hoptimalbaseline}{1}
\newcommand{\hdiffnfc}{2}
\newcommand{\hdiffobjective}{3}

\newcommand{\cmark}{\ding{51}}%
\newcommand{\xmark}{\ding{55}}%

\usepackage{microtype}        % Improved Tracking and Kerning
\usepackage{ccicons}          % Cite your images correctly!
\usepackage{todonotes}

\def\plaintitle{Towards Optimizing Human-Centric Objectives in AI-Assisted Decision-Making With Offline Reinforcement Learning}
\AtBeginDocument{%
  \providecommand\BibTeX{{%
    \normalfont B\kern-0.5em{\scshape i\kern-0.25em b}\kern-0.8em\TeX}}}

\setcopyright{acmcopyright}
\copyrightyear{2024}
\acmYear{2024}
\acmDOI{XXXXXXX.XXXXXXX}

\begin{document}

\title{\plaintitle}

\author{Zana Bu\c cinca}
\orcid{0000-0002-2644-6065}
\affiliation{%
  \institution{Harvard University}
  \city{Boston}
  \state{Massachusetts}
  \country{USA}}
\email{zbucinca@seas.harvard.edu}

\author{Siddharth Swaroop}
\orcid{0009-0009-5345-8844}
\affiliation{%
  \institution{Harvard University}
  \city{Boston}
  \state{Massachusetts}
  \country{USA}}
\email{siddharth@g.harvard.edu}

\author{Amanda E. Paluch}
\orcid{0000-0003-4244-9511}
\affiliation{%
  \institution{University of Massachusetts Amherst}
  \city{Amherst}
  \state{Massachusetts}
  \country{USA}}
\email{apaluch@umass.edu}

\author{Susan A. Murphy}
\orcid{0000-0002-2032-4286}
\affiliation{%
  \institution{Harvard University}
  \city{Boston}
  \state{Massachusetts}
  \country{USA}}
\email{samurphy@g.harvard.edu}

\author{Krzysztof Z. Gajos}
\orcid{0000-0002-1897-9048}
\affiliation{%
  \institution{Harvard University}
  \city{Boston}
  \state{Massachusetts}
  \country{USA}}
\email{kgajos@eecs.harvard.edu}

\renewcommand{\shortauthors}{Bu\c cinca, et al.}

\begin{abstract}

%Background/Motivation
Imagine if AI decision-support tools not only complemented our ability to make accurate decisions, but also improved our skills, boosted collaboration, and elevated the joy we derive from our tasks. Despite the potential to optimize a broad spectrum of such human-centric objectives, the design of current AI tools remains predominantly focused on decision accuracy alone.
%Method
To address this gap, we propose offline reinforcement learning (RL) as a general approach for modeling human-AI decision-making to optimize human-AI interaction for diverse objectives. RL enables optimizing various objectives in AI-assisted decision-making by tailoring and adaptively providing decision support to humans --- the right type of assistance, to the right person, at the right time. We instantiated our approach with two objectives: human-AI accuracy on the decision-making task and human skill improvement (i.e., learning about the task) and learned decision support policies from previous human-AI interaction data.
%Results
We compared the optimized policies against several baselines in AI-assisted decision-making. Across two experiments (N = 316 and N = 964), our results consistently demonstrated that people interacting with policies optimized for accuracy achieve significantly better accuracy --- and even human-AI complementarity --- compared to those interacting with any other type of AI support. Our results further indicated that human learning was more difficult to optimize than accuracy, with participants who interacted with learning-optimized policies showing significant learning improvement only at times. 
%Significance
Our research (1) demonstrates offline RL to be a promising approach to model the dynamics of human-AI decision-making, leading to policies that may optimize human-centric objectives and provide novel insights about the AI-assisted decision-making space, and (2) emphasizes the importance of considering human-centric objectives beyond decision accuracy in AI-assisted decision-making, opening up the novel research challenge of optimizing human-AI interaction for such objectives.

\end{abstract}




\begin{teaserfigure}
\centering
 \includegraphics[scale=0.15]{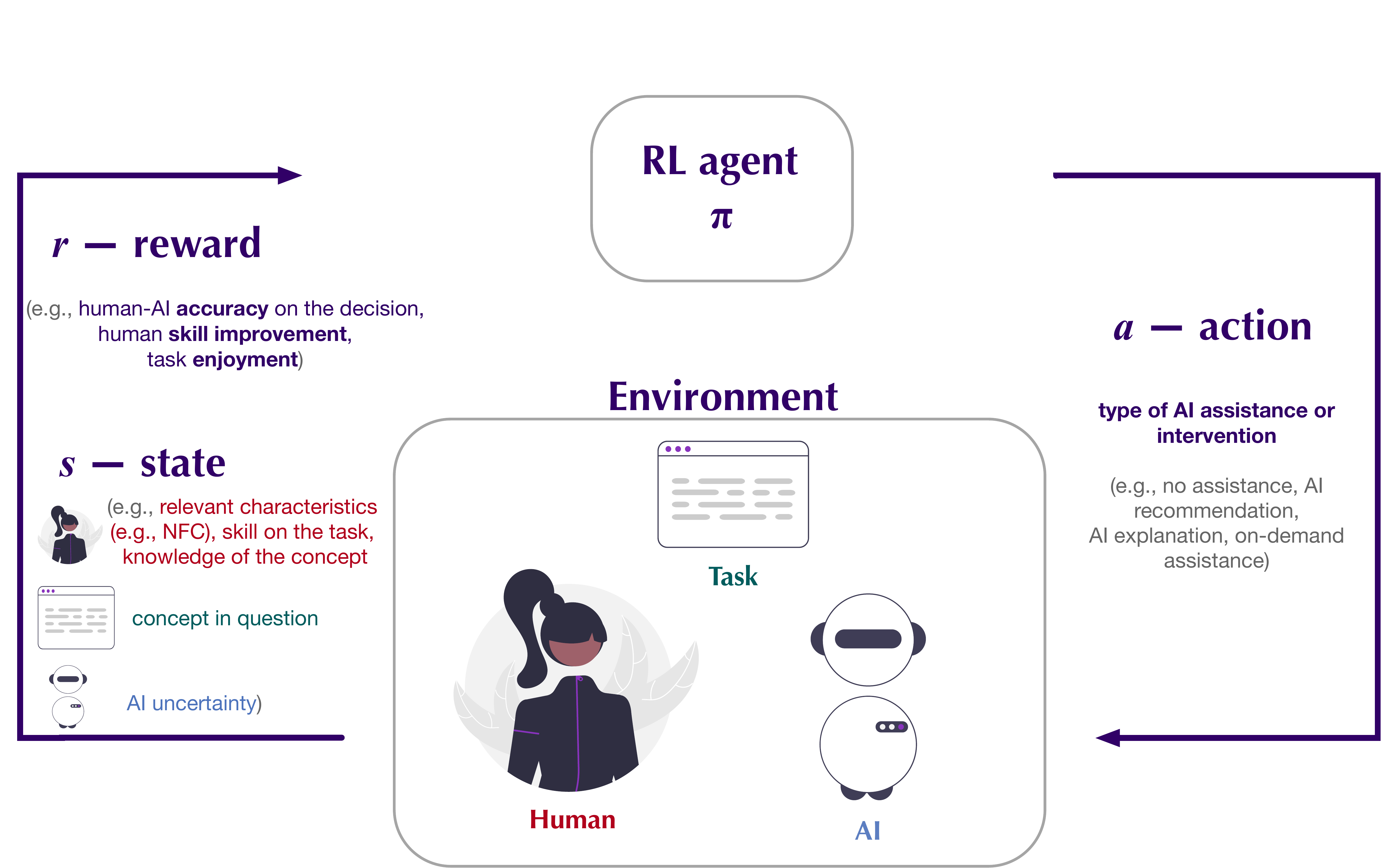}
 \caption{A simplified overview of the proposed method. Providing adaptive decision support through reinforcement learning for optimizing human-centric objectives while accounting for human-centric and other contextual factors in human-AI decision-making.}
 \label{fig:teaser}
\end{teaserfigure}


\maketitle

\section{Introduction}

%multiple objectives
%% -- multiple human-AI interactions developed, with varying effectiveness on the objectives
%As we increasingly make decisions and complete tasks with the assistance of AI-powered technology, how will the quality of our work, our efficiency, our skills, the joy we derive from our tasks, or our collaboration with others be affected?
AI-powered assistance is increasingly woven into our decision-making with the promise of improving the quality and efficiency of our decisions.  However, beyond decision quality and efficiency, how will this assistance affect us --- our skills and growth, the joy we derive from our tasks,  the way we collaborate with others, or the agency we feel in the workplace? The emerging human-AI dyads form sociotechnical systems that produce both tangible (e.g., decisions) and socio-psychological outcomes~\cite{walker2008review, ropohl1999philosophy}. Decades of research in work design have highlighted and empirically demonstrated that human-centric socio-psychological outcomes, like competence or autonomy, are crucial mediators of motivation, performance, and overall well-being in the workplace~\cite{hackman1976motivation, morgeson2003work, parker2024smart, deci2017self, parker2017one}. Yet, the present design and development of AI decision aids has been narrowly fixated on improving only the accuracy of the decisions, largely neglecting other human-centric objectives that the decision-maker may value and find motivating in their work (e.g., skill improvement, autonomy, social belonging~\cite{deci2012self}). While immense effort has been invested in the development of human-AI interaction techniques to enhance accuracy (e.g.,~\cite{miller2023evaluative, danry2023don, bucinca2021trust, mozannar2024effective}), almost no research has considered or explicitly studied how to optimize human-centric objectives in AI-assisted decision-making. We posit that as AI assistance becomes integral to our daily tasks and workflow, it is essential that we maintain control over its both short- and long-term impact on decision-maker's well-being and quality of life. To guide this impact effectively, we must account for the human decision-maker and devise methods and techniques to explicitly optimize for human-centric objectives along with accuracy in AI-assisted decision-making.

%evidence that dynamic support is a good approach to optimizing human-centric objectives
%% -- a flavor of the policies that were learned
%But how to design decision-support that optimizes human-centric objectives?  Empirical evidence from accuracy as the sole objective that has been studied extensively may inform the paradigm of support that may be needed for optimizing human-centric objectives. 
But why may the existing decision support paradigm be insufficient in optimizing such human-centric objectives? First, even with decision accuracy as the primary objective, the design of current AI decision-support tools generally provides a fixed type of support --- like an AI recommendation and an explanation --- regardless of the specific decision, person, or context. Meanwhile, alternative human-AI interaction approaches are being developed (e.g., cognitive forcing~\cite{bucinca2021trust}, evaluative AI~\cite{miller2023evaluative}, self explanation~\cite{danry2023don}, explanation without decision recommendation~\cite{gajos2022people}). Additionally, there is growing evidence that the choice of the optimal human-AI interaction approach ---  even when considering only accuracy as the objective --- depends on multiple contextual factors, including factors specific to a decision instance (such as a person's knowledge and confidence related to particular a decision instance) as well as AI's confidence~\cite{noti2022learning}. However, we do not yet have fully systematized knowledge to indicate what type of interaction should be presented under what circumstances. Considering multiple objectives and importance of context in AI-assisted decision-making, we believe that AI support needs to be dynamic, changing in response to context and individuals while optimizing decision accuracy and other specified human-centric objectives. For instance, such dynamic assistance may prevent human overreliance on AI by withholding AI assistance in cases when the AI is uncertain, rather than offering recommendations and explanations for every decision task. 
When focusing on improving human skills, it may show partial support (e.g., only explanations) instead of providing decision recommendations (``the answer'') to encourage deeper cognitive engagement and learning~\cite{gajos2022people}.  Or envisioning futures in which we seek to enhance collaboration and relatedness in the workplace~\cite{deci2017self}, such dynamic assistance may even at appropriate times advise decision-makers to seek insights from a more experienced colleague, or to form a team to tackle complex decision-making scenarios that an individual is uncertain about.

%Empirical evidence from accuracy as the central objective that has been studied in AI-assisted decision-making strongly suggests that effective AI decision support needs to be dynamic. Yet, the design of current AI decision-support tools generally provide a fixed type of support -- like an AI recommendation and an explanation -- regardless of the specific decision, person, or broader context. Meanwhile, alternative human-AI interaction approaches are being developed (e.g., cognitive forcing~\cite{bucinca2021trust}, evaluative AI~\cite{miller2023evaluative}, self explanation~\cite{danry2023don}, explanation without decision recommendation~\cite{gajos2022people}), and there is growing evidence that the choice of the optimal human-AI interaction approach depends on multiple contextual factors, including factors specific to a decision instance (such as a person's knowledge and confidence related to particular a decision instance) as well as AI's confidence~\cite{noti2022learning}. However, we do not yet have fully systematized knowledge to indicate what type of interaction should be presented under what circumstances.

%% Our proposed approach

To enable dynamic AI support that adapts based on the context and objectives, in this paper, we cast the problem of human-AI decision-making as a Markov Decision Process and learn policies that optimize different objectives with offline reinforcement learning (RL). We propose RL as a particularly appealing approach for optimizing human-centric objectives in AI-assisted decision-making due to its ability to model objectives that are sparse or harder to capture (like human learning or task enjoyment) as part of the reward, to capture human-centric and contextual factors (e.g., human's skill, load, motivation, AI's uncertainty) as part of the state space, and adapt the support with an action space comprised of different types of human-AI interaction techniques effective for specific contexts. Our proposed approach uses \emph{offline} RL to derive optimal support policies from existing datasets of human-AI decisions with various AI assistance. %Offline RL is a safer alternative to real-time exploration, avoiding potentially significant risks and high costs of real-time exploration in actual human-AI decision-making settings (e.g., clinical setting). 
Our proposed approach is highly customizable, with the potential to optimize various objectives in AI-assisted decision-making by appropriately crafting the reward and constructing the state and action spaces. For this paper, we focus on optimizing \em decision accuracy \em and \em human learning \em about the task as two important objectives to optimize in AI-assisted decision-making, and examples of dense (accuracy) and sparse (human learning) rewards  in our proposed approach.  Drawing on insights from prior work, we construct (i) an action space with four different assistance types that may be effective to optimize these two objectives  and (ii) a state space that, along with relevant contextual factors (e.g., AI uncertainty), includes human-centric factors and individual differences, such as people's Need for Cognition (NFC). NFC, a stable personality trait indicating a person's inclination towards cognitively demanding activities, plays a crucial role in determining how likely a person is to engage with AI support~\cite{gajos2022people, bucinca2021trust}, thereby influencing the effectiveness of different human-AI interaction techniques to achieve objectives that require cognitive engagement (e.g., human learning).

%% The instantiation of the proposed approach

%% Key results
We first conducted a data collection study (N=142) in which participants made sequential decisions related to an exercise prescription task. In the data collection study, participants interacted with an exploratory decision-support policy that sampled AI assistance types uniformly.
From these interaction data, we applied Q-learning to learn policies that optimized accuracy (immediate accuracy on the task), human learning (accuracy on post-intervention questions), or a combination of both.
We conducted two types of evaluations of our approach: a computational evaluation of the optimized policies, and two studies with human-subjects interacting with the optimized policies and various baselines. First, our computational analysis and interpretation of the learned policies revealed that optimal policies are different, in meaningful ways and in line with current understanding of the space, for different objectives, contexts, and people with different levels of Need for Cognition (NFC). Examining the policies further led to discovering new insights about the AI-assisted decision-making space. Specifically, we discovered that participants low in NFC are unlikely to request AI assistance when that assistance is offered on demand.
Second, our results from the first (N=316) and second (N=964) human-subject studies demonstrated that policies optimized for accuracy consistently, and significantly improved the immediate decision accuracy of both NFC groups compared to various baselines and learning-optimized policies. On the other hand, the learning-optimized policy led to significantly more learning than the accuracy-optimized policy only for the group low in NFC and only in the first experiment. Our results indicate that it was more challenging to optimize for learning compared to immediate decision accuracy. We believe this is partially due to the weakness of the signal and the lack of actions (i.e., human-AI interaction techniques) that robustly improve learning.

%% Implications

In summary, this paper makes the following contributions:

 \begin{itemize}
        
     \item We introduce and illustrate the potential of offline RL as a promising method for modeling the complexities of human-AI decision-making, allowing the development of decision-support policies that optimize multiple human-centric objectives in AI assisted decision-making.
     \item Our instantiation of the proposed approach is consistently successful in improving joint human-AI accuracy, achieving even human-AI complementarity, but only partially successful in improving human learning.
    \item We further demonstrate the potential of offline RL as a means to discover insights about the AI-assisted decision-making space.
    \item We contribute new evidence demonstrating the significance of individual differences in cognitive motivation (i.e., Need for Cognition), as a factor to be taken into account when designing AI systems for decision support.
    \item Our work opens up a novel research challenge of designing novel paradigms, explanations, and human-AI interaction techniques that optimize learning and other human-centric objectives along with decision accuracy in AI-assisted decision-making.
     %\item We contribute new evidence demonstrating the significance of individual differences in cognitive motivation (i.e., Need for Cognition), as a factor to be taken into account when designing AI systems for decision support.
    %\item We show that overreliance on AI does not necessarily indicate a lack of cognitive engagement.

\end{itemize}
\section{Background \& Related Work}

%%%
% (Beyond Expertise and Rules) Learn about a domain. Through interview studies, both
%Hong et al. [58] and Liao et al. [76] describe how stakeholders across diferent domains use interpretability to generate
%new hypotheses or insights about a domain. For example,
%one participant aimed to use a model predicting surgeons’
%future performances as a tool to better understand what
%factors drive good performance, rather than using it as a
%predictive system. Hohman et al. [56] focus specifcally on
%data scientists, describing how interpretability helped them
%fnd “valuable nuggets of information” in the data. Similarly,
%Doshi-Velez and Kim [38]

%Learn from the user (cite guidelines Saleema)

\subsection{Human-AI Accuracy in AI-assisted Decision-making}

\subsubsection{Towards Calibrated Reliance on AI in AI-Assisted Decision-Making}

AI is becoming increasingly integrated into decision-making processes, with the assumption that it will enhance decision-makers' abilities by combining their expertise with AI advice to improve decision outcomes. However, mounting evidence shows that decision-makers struggle to incorporate AI recommendations into their decisions, often either over-relying or under-relying on AI, even when explanations are provided~\cite{bansal2021does, bucinca2021trust, passi2022overreliance, gaube2021ai, schaffer2019can, chen2023understanding, yin2019understanding, park2019slow, lu2021human, alufaisan2021does, schaffer2019can}.

Recognizing this challenge, substantial efforts have been made to characterize the types of explanations or indicators of uncertainty ~\cite{vasconcelos2023explanations, chen2023understanding, zhang2020effect, yin2019understanding, rechkemmer2022confidence, lu2024does, kahr2023seems}, situations (e.g., the cost-benefit of engaging with AI~\cite{vasconcelos2023explanations}, time pressure~\cite{swaroop2024accuracy, cao2023time}) and other settings~\cite{poursabzi2018manipulating, guerdan2023ground, langer2022look} in which people resort to over- or under-relying on AI and devise interventions that promote calibrated reliance and effective utilization of AI support. These research endeavors can be broadly categorized into pre-task and in-the-moment interventions. Pre-task interventions often involve training or onboarding sessions designed to help individuals construct a mental model of AI~\cite{mozannar2022teaching, mozannar2024effective, pinski2023ai, kawakami2023training}, develop a self-mental model related to the task~\cite{he2023knowing}, or increase human agency by granting them control over input feature selection and algorithmic assistance~\cite{cheng2023overcoming, lai2023selective}. In-the-moment interventions, on the other hand, consist of interventions such as explanation~\cite{yang2023harnessing}, interaction~\cite{bucinca2021trust}, meta-information~\cite{cabrera2023improving, mozannar2024effective}, and paradigms~\cite{miller2023evaluative} that promote effective AI support use during the decision-making process. Some of these interventions can be broadly grouped into evaluation-soliciting decision support, such as Miller's proposed Evaluative AI paradigm, which presents evidence both for and against a decision \em after \em the human makes an initial decision~\cite{miller2023evaluative}. Other approaches involve presenting explanations in the form of questions rather than statements~\cite{danry2023don}, or decision support that incorporates evidence from the literature and presents it alongside AI advice~\cite{yang2023harnessing}.
Related to our work, a nascent branch of in-the-moment interventions includes adaptive strategies that learn to present decision-makers with AI support only when it is deemed beneficial. To identify such instances, these adaptive interventions leverage a model of human decision-makers~\cite{ma2023should, noti2022learning} or learn decision policies with contextual bandits~\cite{bhatt2023learning}. These strategies, which optimize whether or not to provide support for optimizing immediate accuracy, seem promising for effective AI-assisted decision-making: both Noti and Chen~\cite{noti2022learning} and Ma et al.~\cite{ma2023should} report human-AI complementary team performance.

\subsubsection{The Impact of Different AI Assistance on (Over)Reliance and Cognitive Engagement}

Studies consistently show that simple explainable AI (SXAI), in which people are provided with AI recommendations and explanations, induces overreliance on AI~\cite{bucinca2021trust, gaube2021ai, bansal2021does}. Previous research in AI-assisted decision-making has put forth that this overreliance on AI stems from superficial engagement with the information provided ~\cite{bucinca20:proxy, bucinca2021trust, gajos2022people}. People overrely on AI recommendations as they fail to cognitively engage with the presented AI suggestion and explanation. Research from learning sciences has long established that cognitive engagement with information is essential for learning~\cite{rotgans2011cognitive}. As such, AI assistance types that induce overreliance will potentially hurt cognitive engagement and subsequently learning. Whereas AI assistance types that induce cognitive engagement will help people critically evaluate information and disregard incorrect AI suggestions, resulting in both increased learning and reduced overreliance. 

One assistance type that previous work suggests enhances cognitive engagement is providing people with AI explanation only~\cite{gajos2022people}. The underlying hypothesis is that providing people with AI explanations only, as opposed to showing them AI recommendations and explanations, invokes more cognitive engagement because people have to make the cognitive jump of getting to a final decision from the given information rather than being ``served the answer''. 
Another form of AI support that previous work has shown to reduce overreliance, and possibly induce cognitive engagement, is letting people choose whether or when they want to see AI recommendations and explanations (i.e., on demand)~\cite{bucinca2021trust}. By tapping into people's curiosity for viewing the AI advice and allowing them control over when or whether to view the AI suggestion, such assistance may elicit cognitive engagement with the AI-provided content.
We included these two designs in our study, as assistance types that had the potential to support cognitive engagement, and thus, human learning about the domain.

\subsection{Human Competence, AI, and the Future of Work}

The anticipated large-scale deployment of AI-powered decision aids is likely to transform many jobs. There are moral and economic reasons to look for ways to deploy these technologies in a manner that complements workers and enhances their abilities, rather than diminish their roles or replace them without offering new opportunities~\cite{acemoglu2023can, siddarth2021ai}. However, many of the current deployments of AI-powered decision support systems are likely to negatively alter the existing workplace dynamics. For example, when workers rely on one another for help with difficult decisions, such help typically results in \em incidental learning \em that enables workers to develop their skill over time~\cite{berlin1992consultants}. In fact, some researchers argue that a large fraction of learning that occurs in organizations happens via informal channels such as incidental learning~\cite{marsick01:informal,marsick2017rethinking} and such learning is essential not just for workforce development but also for worker well-being~\cite{deci2017self}. However, receiving help from systems that offer a decision recommendation accompanied by an explanation does not seem to result in incidental learning~\cite{gajos2022people}. Instead, it can lead workers to incorrectly increase their confidence in their ability to perform similar tasks in the future~\cite{fisher2021knows,fisher2021harder}.

Further, as AI increasingly assists knowledge workers in decision-making by providing decision recommendations, a critical question arises: how will such assistance affect decision-makers’ work motivation in the long term?
Self-determination theory (SDT) ---  a macro theory for understanding human motivation --- may provide insights into how AI support systems may affect decision-makers' long-term motivation in the workplace~\cite{deci2017self}. Originating from studies on intrinsic and extrinsic motivations and subsequently broadening its scope to encompass investigations in areas such as work organizations and various aspects of daily life, SDT identified \em competence \em as one of the three psychological needs which mediated workers' performance and well-being in the workplace. Competence reflects a person's drive to be effective and skilled in their work environment. It involves exploring and engaging with surroundings and taking on challenging tasks to assess and improve abilities. 
Meanwhile, the current design of AI support with recommendation and explanation (SXAI) may be inadvertently undermining decision-makers' competence. While the longitudinal impact that SXAI has on competence is yet to be studied, evidence from automation~\cite{crocoll1990status} and a recent study in AI-assisted decision-making~\cite{gajos2022people} suggest that decision recommendations may hinder decision-makers' learning and skill improvement.
Given the critical standing of competence in workers' motivation, well-being, and performance, we posit that supporting  workers' skill improvement and knowledge acquisition (along with their decision accuracy) is a critical human-centric objective for the design of AI for decision support.
While reinforcement learning has been applied in educational settings to learn personalized curricula~\cite{doroudi2019s}, to the best of our knowledge, our work is the first endeavor to learn interactive policies that optimize decision-makers' learning in sequential decision-making tasks.

\subsection{Need for Cognition}
Need for Cognition (NFC) is a stable personality trait that captures how likely a person is willing to engage in non-required cognitively demanding activities~\cite{cacioppo82:need}. In other words, it reflects a person's general cognitive motivation. Across, numerous fields such as skill acquisition, processing of information in advertising and in health communication, or web usage, there is consistent evidence that high need for cognition is associated with seeking out more information and processing that information more deeply~\cite{cazan14:need,lin11:effects,vidrine07:construction,williams-piehota03:matching,tuten01:social,sicilia05:effects}. In HCI literature, there is also initial evidence that people high in NFC are more likely than those with low NFC to exert cognitive effort when interacting with complex digital systems and to benefit more from the more complex features~\cite{carenini01:analysis,gajos2017influence,tuten01:social}. In the area of AI-supported decision-making, previous work found that compared to individuals with low NFC, those with high NFC make better decisions~\cite{bucinca2021trust} and benefit more from novel human-AI interaction techniques such as cognitive forcing~\cite{bucinca2021trust} or receiving only explanations without decision recommendations~\cite{gajos2017influence}. For those reasons, we identified need for cognition as a particularly relevant dimension of individual differences.

\subsection{Offline Reinforcement Learning}

Reinforcement learning (RL) is a popular approach to designing intelligent systems that learns by interacting with an environment, and can be divided into two categories based on the data collection strategy: online and offline RL~\cite{sutton2018reinforcement}. Online RL entails learning optimal policies through direct interaction with the environment, either in the real world or within a simulated setting, in real-time. Whereas, offline RL involves learning optimal policies from a previously-collected interaction dataset. One of the main advantages of online RL is its ability to adapt to changes in the environment and to learn in real time, making it well-suited for applications that require continuous adaptation. Online RL has been employed in various applications of user interfaces (e.g., menu selection~\cite{li2023modeling}, interface adaptation~\cite{todi2021adapting}, visual search~\cite{chen2017cognitive}, typing~\cite{jokinen2021touchscreen}) in which policies are typically learned by interacting with a computational model or simulation of the user behavior.
Online RL with real users (i.e., in a real-world environment) can be risky (due to exploratory actions taken in real time), computationally expensive (especially in time-constrained settings), and data-intensive, with data collection in the real world often being costly.
Offline RL, on the other hand, is safer and less computationally expensive, as it learns from a fixed dataset before the policy is deployed. However, offline RL may not generalize well to new environments (such as when parts of the environment are not sufficiently explored in the offline training dataset)~\cite{levine2020offline}.
We opted learning policies from actual human-AI interaction data as opposed to simulations. With faithful computational models or simulators of human-AI interaction, learning policies with simulated data would also be possible. However, our understanding of human-AI decision-making as a field is still in its infancy, therefore, any assumptions baked into a computational model may likely turn to be incorrect, and thus, yield flawed policies (e.g., One such assumption was that explanations would help people calibrate their reliance on AI; they generally do not~\cite{bansal2021does}.). Because of the real-world constraints and cost of employing online RL for crowdsourcing studies with actual users, we chose an offline RL setup and collected data accordingly. We learned our policies using Q-Learning on data that was previously collected using an exploratory policy. 

Q-Learning~\cite{watkins1989learning} is a prominent off-policy algorithm in RL for data collected in an offline setting. It estimates the expected long-term reward of each state-action pair (the ``Q-value'') in the RL environment.
%Off-policy algorithms like Q-learning are a popular approach to RL that learn a value function that estimates the expected long-term reward of a given action in a given state. 
Q-learning estimates a state-action pair's Q-value by iteratively updating it based on the observed reward and the estimated value of the next state.
Once Q-learning converges, the optimal policy is the action with the highest value for each state. One of the key advantages of off-policy algorithms is that they can learn from data collected using any policy, which allows the agent to learn from potentially suboptimal but diverse behaviors. For example, collecting data using an exploratory policy can help to prevent the agent from getting stuck in suboptimal behavior and to learn from potentially rare but valuable experiences.

\section{Approach, Setting \& Problem Formulation}

\subsection{Approach}

In this work, we sought to build computational models that dynamically select interactions based on (1) the desired objectives of AI-assisted decision-making, (2) the individual differences among the decision-makers, and (3) the relevant contextual factors. For this paper, the two objectives we chose to optimize for were people's immediate decision accuracy and their longer-term learning about the task (as measured by their accuracy on distal tasks where they receive no AI support). The individual difference as a relevant human-centric factor for which we personalized was people's Need For Cognition (low vs. high), drawing on results from prior work that have shown it to be an important predictor of engagement with different forms of AI support~\cite{bucinca2021trust, gajos2022people}. Informed by previous work discussed above, the contextual factors we considered were the type of decision-making instance, AI's uncertainty, decision-maker's competence, and knowledge related to the concept in question. 

Our setting was a sequential decision-making task, in which the same individual made a large number of decisions in the presence of an AI-powered decision support system. On any particular decision task instance, the decision support system could support the human decision-maker using one of several methods (no support, showing a decision recommendation with an explanation, showing just an explanation but offering no decision recommendation, or allowing the person to request support on demand), each with a different possible impact on a person's decision performance in the moment and on their learning about the task domain. We formulated the problem of personalizing AI support as a Markov Decision Process, employing reinforcement learning (RL) algorithms to learn optimal policies that select sequences of interactions by accounting for the context and individual differences when optimizing the desired objectives (See Section~\ref{section:problem-formulation}).

Our work was guided by the following three broader hypotheses about adaptive support in AI-assisted decision-making which we further broke down into specific hypotheses in subsequent sections:

\begin{enumerate}

    \item Optimized policies will result in human performance on the target objectives (i.e., accuracy and learning) that is as good or better than the performance achieved with baseline policies that do not consider contextual factors.
    
    \item People with different levels of Need for Cognition (NFC) will benefit from different types of AI assistance for objectives that require cognitive engagement (i.e., learning). 

    \item For each group of people based on their level of NFC, policies optimized for a target objective (i.e., accuracy or learning) will result in better human performance on that objective than policies optimized for another objective.
\end{enumerate}

% \section{Decision-making Task: Exercise Prescription}
\subsection{Setting: Exercise Prescription Decision-Making Task}

\begin{figure*}
     \centering
     \begin{subfigure}{\textwidth}
         \centering
         \includegraphics[width=\textwidth]{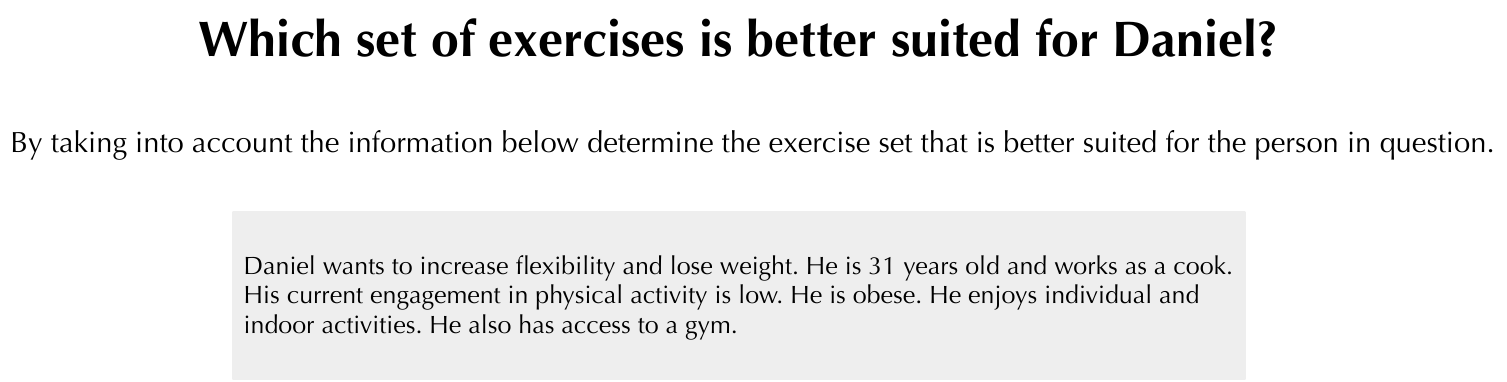}
         \caption{Stimulus}
         \label{fig:stimulus}
     \end{subfigure}
     \hfill

     \begin{subfigure}{0.3\textwidth}
         \centering
         \includegraphics[width=\textwidth]{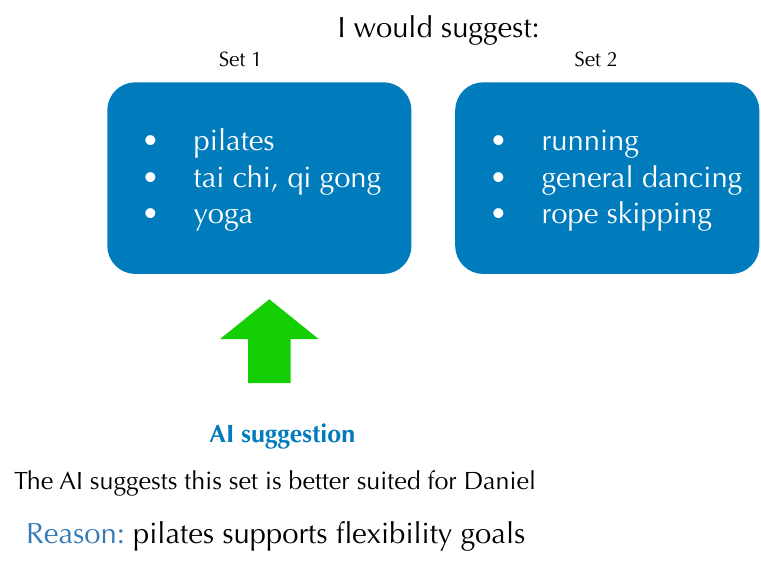}
         \caption{Recommendation and explanation}
         \label{fig:rec_explanation}
     \end{subfigure}
     \hspace{15px}
     \begin{subfigure}{0.3\textwidth}
         \centering
         \includegraphics[width=\textwidth]{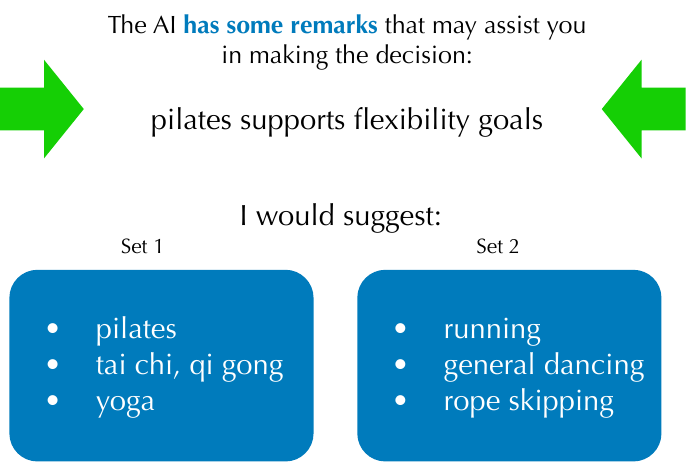}
         \caption{Explanation only}
         \label{fig:explanation}
     \end{subfigure}
     \hspace{15px}
     \begin{subfigure}{0.3\textwidth}
         \centering
         \includegraphics[width=\textwidth]{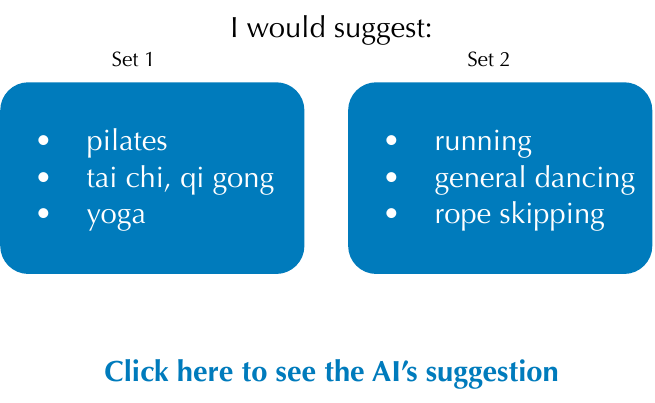}
         \caption{On demand}
         \label{fig:on_demand}
     \end{subfigure}
     \caption{An example of the exercise prescription decision-making task with different types of AI assistance (i.e., actions). Participants were assisted in choosing between the two sets of exercises as depicted for different conditions. In the \em No-AI \em condition (not shown) participants were not provided with any AI assistance.} 
     \label{fig:exercise-task}
\end{figure*}

Figure~\ref{fig:exercise-task} depicts an example of the decision-making task. We aimed to create a decision-making task that would be accessible to laypeople on crowd-sourcing platforms, while also inducing similar cognitive challenges as high-stakes decision-making tasks such as those encountered in clinical decision-making. To accomplish this, we teamed up with a kinesiology expert (also a co-author of this paper) and created decision instances for an exercise recommendation task. The task involves selecting the most suitable of two exercise sets for a (fictional) person based on their description, goals, and preferences. Exercise recommendation as a decision-making task is accessible to a broad audience, yet poses similar challenges to treatment selection in clinical settings. When choosing a treatment, clinicians weigh various factors such as the patient's condition, treatment preferences, side-effect tolerance, and constraints. Similarly, in exercise prescription, each exercise type may interact differently with the person's goals, health factors, preferences, capabilities, or resources.

We generated 44 vignettes of fictitious people by randomly sampling their demographics (age, gender, BMI, physical activity level, occupation, smoking status) and accordingly manipulating the following six factors which were deemed important for exercise prescription by the expert: (1) their maximal or target intensity (based on demographics), (2) their exercise goal (\em e.g.\em, building muscles, weight loss, flexibility), (3) their exercise preference (\em e.g.\em, indoor/outdoor, group/individual), (4) their resource availability (\em e.g.\em, access to a swimming pool), (5) their medical condition if any (based on their age and gender), and (6) their susceptibility to experience adverse events during the exercise (based on their age and medical condition).

To build an exercise repository from which to recommend exercises to the fictitious people, we curated a list of 60 leisure time exercises from a comprehensive compendium which consisted of physical activities ranging from different exercises (\em e.g.\em, sports) to everyday activities (\em e.g.\em, housework, occupational activities)~\cite{ainsworth20112011}.
Given a fictitious person and the list of exercises, the expert selected a list of \em optimal \em and a \em suboptimal \em exercises for the person. The \em optimal \em and \em suboptimal \em choices differ substantially in at least one of the four concepts (intensity, goal, medical condition, and safety), which rendered the optimal exercise set choice superior to the suboptimal choice. For example, for a fictitious person whose goal is to increase flexibility, optimal exercises such as \em pilates \em and \em yoga \em would account for their goal, while suboptimal choices such as \em running \em or \em rope skipping \em may fit their other needs but would not support their goal.

\textbf{Generating explanations.} In order to generate effective explanations, we carefully crafted them to highlight the specific concept of one of the exercises that resulted in one exercise set being more optimal than the other. Thus, if one set of exercises was superior to the other due to the medical condition, the explanation would highlight the feature of one of the exercises in the superior set that makes it suitable for the person in question. For example, if the person in question had osteoporosis and the superior set included low-impact exercises such as \em swimming\em, the explanation would take the form: \textit{swimming is suited for people with osteoporosis because it is low-impact.}

\textbf{Generating incorrect AI recommendations.} Our experiment aimed to recreate realistic scenarios in which an AI model assisting individuals during decision-making might not always be accurate. To simulate such situations, we used a subset of the questions (25\% for the data collection study and 28.6\% for the evaluation study)  in which the AI assistance presented the suboptimal choice as the recommendation and generated a correct but unhelpful explanation. The explanation highlighted an arbitrary concept for which the suboptimal exercise was superior to the optimal exercise. This approach allowed us to test participants' ability to recognize the suboptimal concept and make a sound decision despite the explanation provided. For example, for the same fictitious person as before whose goal was to increase flexibility, optimal exercises such as \em pilates \em and \em yoga \em would account for their goal, while suboptimal choices such as \em running \em or \em rope skipping \em may fit their other needs but would not support their goal. An unhelpful explanation of the incorrect AI suggestion would be: \textit{running maximizes the intensity the person is capable of exerting}.

\subsection{Problem Formulation}

\label{section:problem-formulation}

Our goal was to optimize for accuracy on the current task instance as a proximal outcome (dense reward), and human learning about the task domain as a distal outcome (sparse reward). To achieve this goal, we formulated the problem of choosing the decision support method for a specific task instance as a reinforcement learning problem.

Let the Markov Decision Process (MDP) be defined by the tuple $(\mathcal{S}, \mathcal{A}, \mathcal{T}, \mathcal{R}, \gamma)$ where $\mathcal{S}$ is the state space, $\mathcal{A}$ the action space, $\mathcal{T}: \mathcal{S} \times \mathcal{A} \to \Delta(\mathcal{S})$ is the probability transition function, $\mathcal{R}$ is the reward, and $\gamma$ corresponds to the discount factor. A policy $\pi: \mathcal{S} \times \mathcal{A} \to [0, 1]$ assigns each state $s \in \mathcal{S}$ a distribution over actions $\pi(a|s)$, where $a \in \mathcal{A}$. In our setting, $\mathcal{S}, \mathcal{A}, \mathcal{R}$ are designed as follows: 

\begin{itemize}

    \item \textbf{State.} In order to capture the current state of the human-AI decision-making dyad, we consider the concept under investigation, AI's accuracy, and the decision-maker's level of knowledge and their propensity to engage in analytical thinking. Specifically, at each time step $t$, we represent the state as $s_t = [{nfc}_t, c_t, u_t, h_t, k_t]$, where:
    \begin{itemize}
        \item ${nfc}_t$ denotes the decision-maker's Need for Cognition (NFC)~\cite{cacioppo1982need}, a personality variable that measures intrinsic motivation to think. To capture this, we asked participants to answer four questions with the highest factor loading from the NFC questionnaire~\cite{cacioppo1982need}, and then categorized their score as \em low \em if it fell below the median or \em high \em otherwise.
        \item $c_t$ represents the concept that is being queried at time step $t$, which makes the optimal exercise superior to the suboptimal exercise for the given vignette. For example, if the task is to choose between walking and ice-skating, and the relevant factor that tips the balance in favor of walking is safety, then safety is the concept that is being probed. In total, four different concepts may be queried: intensity, goal, safety, and condition.  % \sam{we are using the term "concept" to mean two different things.   Around line 330 or so, there were 6 concepts.  Here there are 4 concepts.  Would be good to use different terms.  Maybe "critical factor" is $c_t$?}
        \item $u_t$ represents the model uncertainty regarding its prediction for the current question, which can be modeled as a continuous variable. However, we utilize a binary variable based on the ground truth of the AI's accuracy (i.e., correct or incorrect). We acknowledge that this simplification is not realistic; nevertheless, it minimizes potential confounding variables and enhances our confidence in the overall findings.
        \item $h_t$ captures the decision-maker's average knowledge of concept $c_t$ up to time step $t-1$. It is calculated as the decision-maker's average accuracy over all previous questions about concept $c_t$, discretized to a binary variable with a threshold of 0.6. %$h_t = \frac{\sum_{i=1}^{t-1} p_i \mathbbm{1}_{c_i = c_t}} {\sum_{i=1}^{t-1} \mathbbm{1}_{c_i = c_t}} \geq 0.6 $  \sam{would be good to provide formula for creating $h_t$}
        \item $k_t$ captures the decision-maker's knowledge about the task, measured by their performance on three initial questions with no assistance at the beginning of the study. We classify their knowledge as either \em low \em or \em high\em, depending on whether their average performance was below or above 0.5.
        
    \end{itemize}

    With four possible values for the concept and two values for each of the other four dimensions, our state space consisted of a total of 64 possible states.
    
    \item \textbf{Action space} is comprised of four different interface presentations.
    \begin{itemize}
        \item No assistance: participants receive no AI assistance.
        \item Explanation only: the AI explanation is shown with no recommendation. Previous work suggests that showing explanations only fosters learning about the domain~\cite{gajos2022people}.
        \item AI explanation and recommendation (SXAI): both AI recommendation and explanation about the decision are shown. Numerous studies have demonstrated such a design to increase accuracy because of (over)reliance on the AI~\cite{bucinca2021trust, bansal2021does, gajos2022people}.  Note that throughout the paper for clarity, we refer to the policy that presents AI recommendation and explanation on each decision as simple explainable AI (SXAI), whereas to the type of assistance as \em recommendation and explanation\em. 
        \item On demand: the AI recommendation and explanation are shown upon request. As this action elicits curiosity about the AI's prediction, we hypothesized that it would increase cognitive engagement with the task~\cite{bucinca2021trust}, thus learning as well.
        %\item \sam{would be good to tell reader somewhere what fraction of each of the prior 3 actions is an AI giving low quality advice.   I state this because if the fractions between the actions differ then this could influence the comparison between the actions.}
    \end{itemize}
      \item \textbf{Reward} is multi-objective, seeking to maximize a combination of accuracy (whether the answer in the current question is correct) and learning (whether the provided answer in the later test question with no AI assistance is correct). An answer is considered correct when it matches the expert ground truth. We denote the reward as follows: 

      \begin{equation}
      \label{ref: reward}
        r = (1 - \lambda) p + \lambda  d
      \end{equation}

 where $p$ indicates the \em accuracy \em outcome, $d$ indicates the \em learning \em outcome, and $\lambda$ is the weighing hyperparameter over these two outcomes. For learning, we receive the reward only at test time. However,  in offline RL  all the data is accessible, allowing assignment of credit to an action at the current time for its impact on learning that is assessed in the future. In other words, if a concept was presented with an action $k$ at time $t_i$,  we assign to the action $k$ the future learning reward that is measured at test time~\footnote{This approach to assigning reward is an implementation choice we made for this paper and should not be construed as the only or optimal method for such assignment.} $t_j$, where $j > i$.
\end{itemize}

The data of each participant $i$ is considered as an episode:
$$\mathcal{D_i} = \{ s_1, a_1, s_2, a_2, ..., s_t, a_t, ..., s_{T+1}\}$$ where $T$ is constant and presents the length of the experiment (24 questions), $S_t \in \mathcal{S}$, $A_t \in \mathcal{A}$, and $R_t \in \mathcal{R}$ (the reward received at time step $t$).
Note that the underlying transition probabilities $T(s'|s, a)$ are not directly known in model-free offline policy learning. Instead, the agent infers the dynamics of the environment from the observed transitions in the dataset~\cite{levine2020offline}.

We operate under the Markov property assumption (``the future is independent of the past given the present'').
\section{Method}

Having formulated the problem as a reinforcement learning (RL) problem, we sought to collect data from which to learn policies that optimize for the decision-makers' proximal and distal benefits when assisted by an AI. 

\begin{figure*}
    \centering
    \includegraphics[width=0.7\textwidth]{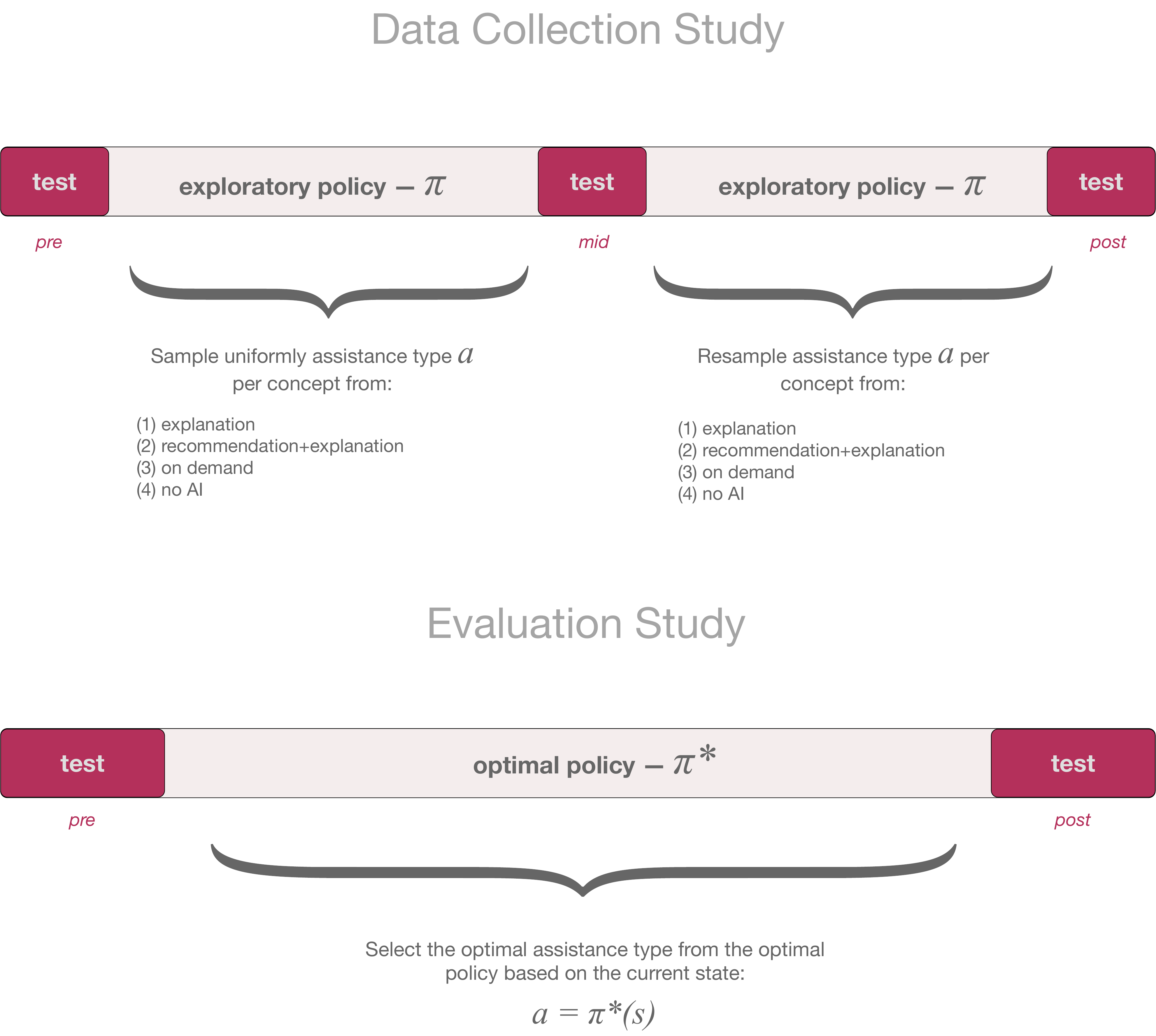}
    \caption{An overview of the experiment flow for the data collection and evaluation studies. In the evaluation studies, participants were randomly assigned to one of the optimal policies (that matched their NFC level) or a baseline policy.}
    \label{fig:experiment-flow}
\end{figure*}

\subsection{Data Collection with an Exploratory Policy}

The purpose of the data collection study was to collect human-AI decision-making data to inform the development of interaction policies aimed at optimizing either immediate decision accuracy or the learning of decision-makers.

\subsubsection{Task}

Participants completed a series of 33 questions, in which each question concerned a vignette about a fictitious character; participants were tasked with selecting which of two sets of exercises was optimal for the character in question.

\subsubsection{Experiment design}

We sought to design an experiment that would enable modeling the impact of various types of assistance on participants' immediate accuracy and their learning about the domain. As such, each participant was randomly assigned to a subset of three different concepts from a pool of four concepts and answered 11 questions per concept, a total of 33 questions. This design choice was informed by work in educational research that highlights the importance of repeated exposure to a concept for effective learning~\cite{koedinger2023astonishing}.

Figure~\ref{fig:experiment-flow} illustrates the experiment design, which consisted of three test blocks (\em pre\em, \em mid\em, and \em post\em) and two intervention blocks (\em first \em and \em second\em). Participants received no AI support on test blocks, which served as evaluation points to measure their initial knowledge and subsequent learning about the domain. On a given intervention block, a concept was presented only with one type of assistance (randomly picked), to isolate the effect of the assistance type on learning. The intervention blocks were preceded and followed by a test block. Notably the \em mid \em test block served as a \em post \em intervention evaluation for the first intervention block and a \em pre \em intervention evaluation for the second intervention block. Each test block consisted of one question per concept, a total of 3 questions. Each intervention block consisted of 4 questions per concept, a total of 12 questions.

The AI system had an overall accuracy of 75\% (three out of 12 intervention questions per block were incorrect). The order of the questions was randomized for each participant and the questions in which the AI made incorrect suggestions were picked randomly. We did, however, ensure that AI was uniformly wrong across concepts, i.e., each of the three incorrect questions per block belonged to a different concept.

In an intervention block, each of the three concepts was quasi-randomly matched with an AI assistance type from the four available AI assistance types --- \em no AI support\em, \em on demand\em, \em explanation\em, \em recommendation+explanation\em. We refer to the procedure as \em quasi \em random because we purposefully sampled \em no AI \em less often than the other forms of AI support. Throughout a given block, a concept was presented only with the assistance type it was assigned to. The same process of assigning concepts to AI assistance types was repeated for the second intervention block. This design choice enabled collecting a larger amount of data per participant to model the impact of different assistance types on immediate accuracy and learning.

\subsubsection{Participants}

%The experiment was a micro-randomized one, as treatments (assistance types) were sequentially randomized throughout the study. 
Given that there are no guidelines for determining the sample size from which RL-based algorithms would reliably capture a signal from the data, the sample size was informed by MRT-SS Calculator, a sample size calculator for micro-randomized trials~\cite{liao2016sample}. A sample of 139 participants is required to attain 80\% power with a significance level of 0.05, and 24 (intervention) decision points (i.e., questions). \footnote{This sample size only reflects binary treatments and focuses on individual treatment effects, not the policies that can be learned from RL algorithms with the given data.}
We recruited a total of 150 participants for the data collection study via Prolific, an online recruitment platform. We retained 142 participants for analysis, contingent on their performance surpassing the attention check threshold. Given that the task involved making decisions based on the comprehension of vignettes, participation was limited to adults in the United States who were fluent in English (for detailed demographics, see Appendix Table~\ref{table:participants}). 
Each participant was compensated 2.4 USD (12.72 USD per hour). %Participants were categorized as \em low \em in NFC if their NFC score fell within the lower 50th percentile, and \em high \em otherwise.

\subsubsection{Procedure}

Our online study was administered through Prolific. Participants were initially provided with a brief overview of the study, and if they agreed to participate, they were directed to an informed consent form. Participants were then required to complete a Need for Cognition (NFC) questionnaire, which included the four items with the highest factor loading from the widely used 18-item instrument~\cite{cacioppo1982need}, as identified in previous work~\cite{gajos2017influence}. Additionally, participants were given the option to complete a demographics form.

Following the completion of these forms, participants were presented with detailed instructions about the task. They were also informed that at times they may receive assistance from an AI that was still under development and was prone to error. The task consisted of answering 33 questions involving exercise recommendation for a series of fictitious characters. At the study's conclusion, participants were asked to report any technical issues they encountered, as well as any instances of cheating. Lastly, participants were provided with feedback on their performance during the study.

\subsubsection{Approvals}
All experiments reported in this paper were approved by our institution's Internal Review Board, under the protocol number IRB21-0805. 
%IRB21-0805.

\subsection{Learning the Optimal Policies}
%number of states

From the collected data, our goal was to learn optimal policies with which to provide assistance to the human decision-maker in order to optimize their accuracy or learning.

We opted for offline reinforcement learning approaches, which aim to learn the optimal policy $\pi^*$ from an exploratory, behavioral policy $\pi_{\beta}$. In our setting $\pi_{\beta}$ is the quasi-uniform policy with which data was collected. We picked Q-learning, as a model-free off-policy RL algorithm that does not require state transition probabilities.

The Q-function was learned by iterating, 
\begin{equation}
    Q(s, a) \gets  Q(s, a) +  \alpha(r + \gamma \max_{a'} Q(s', a')-Q(s,a))
\end{equation}
with the next state in the episode $s'$ (i.e., next question for a participant) and learning rate $\alpha$. We used learning-rate decay on $\alpha$ to speed up convergence, set at a value of $0.1/i$, where $i$ was the number of iterations passed.  In our setting, the reward $r$ is substituted by $(1 - \lambda) p + \lambda  d$ as in equation~\ref{ref: reward}.

\textit{Optimizing accuracy.} When optimizing for accuracy, we myopically seek to maximize the immediate accuracy, disregarding both learning and future rewards. As such, we set both the hyperparameter $\lambda = 0$ and the discount factor $\gamma = 0$. In this way, the algorithm learns to select actions by solely considering the immediate decision accuracy.

\textit{Optimizing learning.} On the other hand, optimizing learning requires consideration of both the learning ($\lambda = 1$) and the future rewards ($\gamma = .99$).

For each objective, we ran 200 iterations over the 142 episodes (i.e., participants), achieving convergence of the \em Q\em-table. Finally, we constructed the optimal policy by picking the optimal action greedily, $\pi^*(s) = \underset{a}{\textrm{argmax}} \, Q(s, a)$ for each state $s$, thereby forming a mapping of states to corresponding optimal actions.

\subsection{Computational Evaluation of the Learned Policies}

First, we examined computationally the learned policies through the lens of the broader hypotheses regarding differences among different objectives and different groups of NFC. Sections~\ref{section:experiment-1} \& ~\ref{section:experiment-2} describe the subsequent evaluation of the policies via two user studies. 

\subsubsection{Hypotheses \& Research Questions}
With the computational evaluation, we sought to answer the following hypotheses and research questions:

\begin{itemize}
    \item \textbf{H\hoptimalbaseline{a}:} Optimized RL policies will differ from fixed policies (e.g., simple explainable AI policy). Specifically, at least one policy for each NFC group will employ a set of interactions that differs significantly from a fixed policy that always uses simple AI recommendation and explanation interactions.
     \item \textbf{H\hdiffnfc{}:} RL policies optimized for improving human learning, as an objective which requires cognitive engagement, will employ different interactions for individuals low in NFC than RL policies optimized for improving the learning of individuals high in NFC.
     \item \textbf{RQ1:} Will RL policies optimized for improving the immediate decision accuracy of individuals low in NFC employ different interactions than RL policies optimized for improving the immediate decision accuracy of individuals high in NFC?
     \item \textbf{RQ2:} Will RL policies optimized for improving the learning of each group of NFC employ different interactions than RL policies optimized for improving immediate decision accuracy?
    
\end{itemize}

\begin{figure*}
    \centering
    
    \includegraphics[width=0.7\textwidth]{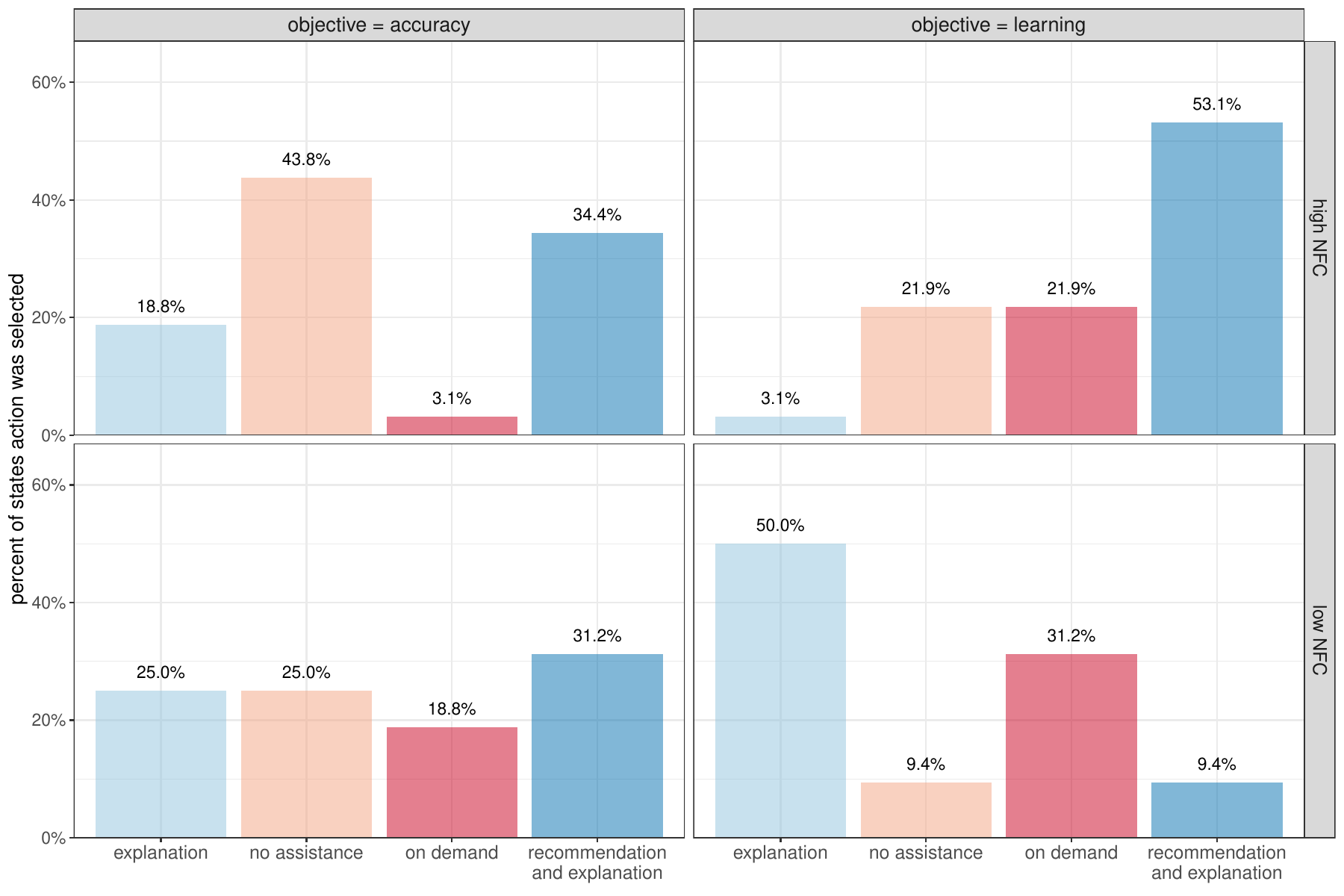}
    \caption{Distributions of types of AI assistance selected by the optimal policies for different objectives and NFC groups. Each bar in the figure represents the percentage of states in which an action was the top action, with the numerator being the number of states where the action was the top choice and the denominator being the total number of states in the analysis.}
    \label{fig:policy-distributions}
\end{figure*}

Figure~\ref{fig:policy-distributions} depicts the distributions of the types of AI assistance for different objectives and different NFC groups. 

\begin{figure}
    \centering
    
    \includegraphics[width=\linewidth]{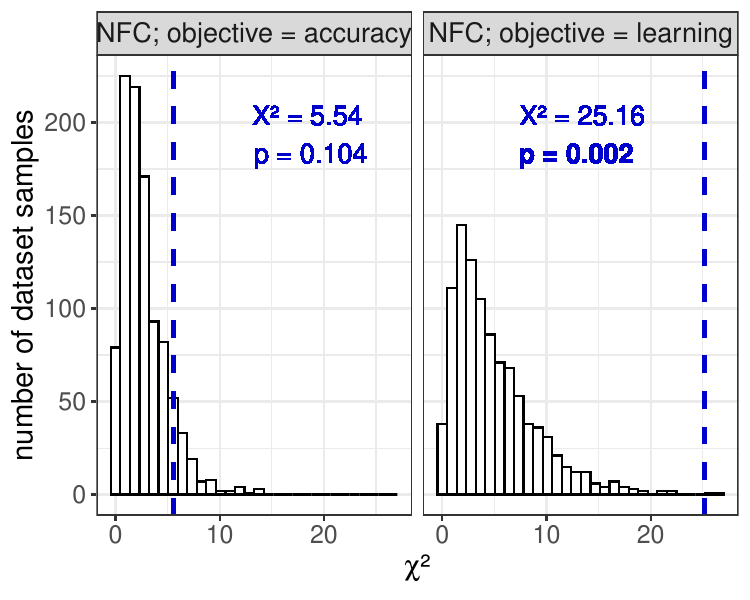}
    \caption{Randomization test results. Each facet depicts the $\chi^2$ distribution of 1000 datasets of random NFC assignments for the given analysis and the $\chi^2$ on the actual dataset (in blue). ``NFC objective = accuracy'', for example, shows the difference of distributions of actions between the two NFC groups for the accuracy as the objective. P-value is computed as the fraction of sampled datasets in which the dataset's $\chi^2$ exceeded the actual $\chi^2$.}
    \label{fig:randomization-test}
\end{figure}

%\subsubsection{People with different levels of Need for Cognition (NFC) will benefit from different types of AI assistance.}

\subsubsection{Results}

%%face validity of the learned policies
In line with \textbf{H\hoptimalbaseline{a}}, both the \em accuracy \em and the \em learning \em policies differ substantially from any fixed policy, such as the SXAI policy, which employs only the recommendation and explanation action. We are unable to conduct a $\chi^2$ test due to its inapplicability for distributions containing zero occurrences, which is the case for all actions other than \em recommendation and explanation \em in the SXAI policy. 

To understand whether different NFC groups benefited from different types of AI assistance, we conducted a randomization test~\cite{imbens2015causal} using $\chi^2$ statistics.  The $\chi^2$ statistic provides a way to compare the action distributions identified as optimal by AI policies for individuals categorized as having low NFC to those with high NFC. 
Because the analysis is conducted over the frequencies of optimal actions for different states that the RL policy has picked, the difference of distributions of actions we obtained might be due to the sample of the participants and not the variable we sought to personalize for --- participant's NFC.  To understand how extreme the obtained $\chi^2$ statistics are, (i.e., whether NFC is indeed the factor that predicts the difference in the distributions of optimal actions) we conducted a randomization test by randomly assigning participants to different NFC groups (regardless of their actual level of NFC). We constructed 1000 such datasets with random NFC assignment. For each of the newly constructed datasets, we learned policies (separately for each of the two objectives) and tested the differences in the distributions of actions via $\chi^2$ (See Figure~\ref{fig:randomization-test}). We report the \em p-value \em, which corresponds to the fraction of the 1000 times that the $\chi^2$ statistic of a dataset with random NFC assignment exceeded the $\chi^2$ statistic of the actual dataset~\cite{imbens2015causal}.

%The use of bootstrapping is used to evaluate if the obtained statistics are unusual across repeated samples of participants\sam{cite Efron's book--see https://www.amazon.com/Introduction-Bootstrap-Monographs-Statistics-Probability/dp/0412042312}. --there should be a chapter on Bootstrap p-values.  You are essentially getting an approximate p-value for the null hypothesis that the true optimal policy is the same regardless of NFC status.}

Supporting \textbf{H\hdiffnfc{}}, our results show that the NFC group is a significant predictor of the distributions of types of assistance for \em learning \em as the objective $(\chi^2(3, N = 128) = 25.16, p = .002)$. Specifically, actions that previous work has shown may elicit cognitive engagement~\cite{bucinca2021trust, gajos2022people} --- \em explanation only \em and \em on demand \em --- were picked more often for people low in NFC than for those high in NFC. 
Whereas for \em immediate accuracy \em as the objective, we do not find any significant difference among the distributions of actions for the two NFC groups $(\chi^2(3, N = 128) = 5.54, p =.10)$, answering \textbf{RQ1}. For optimizing accuracy, actions such as \em no assistance \em (on states where the AI is incorrect) and \em recommendation and explanation \em (on states where the AI is correct) seem to be optimal for both groups.

To determine if the objective -- immediate accuracy or learning -- influences the distribution of AI assistance types (\textbf{RQ2}), using a $\chi^2$ test is inappropriate. This is because the policies for both objectives are derived from the same dataset (i.e., same participants), violating the independence assumption required for the test. However, we observe that when optimizing for accuracy for people high in NFC, \em no assistance \em is selected more often than when optimizing learning. For people low in NFC, both \em no assistance \em and \em recommendation and explanation \em are shown more often when optimizing for accuracy than when optimizing learning.

% \usepackage{multirow}
\begin{table*}[]
\scriptsize
\begin{tabular}{llll}
\hline
                                                                                                                                                                                                                                                         &                                                                                          & \textbf{Experiment 1 (N = 316)}                                                                                                                                                       & \textbf{Experiment 2 (N = 964)}                                                                                                                                                                          \\ \hline
\multicolumn{1}{l|}{\textbf{Hypotheses}}                                                                                                                                                                                                                 & \multicolumn{1}{l|}{\textbf{\begin{tabular}[c]{@{}l@{}}target\\ objective\end{tabular}}} & \multicolumn{1}{l|}{\begin{tabular}[c]{@{}l@{}}baseline: SXAI\\ optimized: accuracy, learning\end{tabular}}                                                                           & \begin{tabular}[c]{@{}l@{}}baselines: SXAI, explanation, random\\ optimized: accuracy, combined, learning\end{tabular}                                                                                   \\ \hline
\multicolumn{1}{l|}{}                                                                                                                                                                                                                                    & \multicolumn{1}{l|}{\textbf{accuracy}}                                                   & \multicolumn{1}{l|}{\begin{tabular}[c]{@{}l@{}}\cmark Supported for both NFC groups\\ \\ accuracy $>$ baseline (SXAI)\end{tabular}}                           & \begin{tabular}[c]{@{}l@{}}\cmark Supported for both NFC groups\\ \\ accuracy $>$ all baselines\\ combined $\ge$ all baselines\end{tabular}                            \\
\multicolumn{1}{l|}{\multirow{-4}{*}{\begin{tabular}[c]{@{}l@{}} \textbf{H1b:}\\ Policies optimized for a target objective will result in human performance on that objective \\ that is either equal to or better than baseline policies.\end{tabular}}}   & \multicolumn{1}{l|}{\cellcolor[HTML]{EFEFEF}\textbf{learning}}   & \multicolumn{1}{l|}{\cellcolor[HTML]{EFEFEF}\begin{tabular}[c]{@{}l@{}}\cmark Supported for both NFC groups\\ \\ learning $\ge$ baseline (SXAI)\end{tabular}} & \cellcolor[HTML]{EFEFEF}\begin{tabular}[c]{@{}l@{}}\cmark Supported for both NFC groups\\ \\ learning $\ge$ all baselines\\ combined $\ge$ all baselines\end{tabular} \\ \hline
\multicolumn{1}{l|}{}                                                                                                                                                                                                                                    & \multicolumn{1}{l|}{\textbf{accuracy}}                                                   & \multicolumn{1}{l|}{\begin{tabular}[c]{@{}l@{}}\cmark* Supported for people high in NFC\\ \\ accuracy $>$ learning\end{tabular}}                              & \begin{tabular}[c]{@{}l@{}}\cmark Supported for both NFC groups\\ \\ accuracy $>$ learning\end{tabular}                                                                          \\
\multicolumn{1}{l|}{\multirow{-4}{*}{\begin{tabular}[c]{@{}l@{}}\textbf{H3a \& H3b:}\\ \\ Policies optimized for a target objective will result in better human performance on\\ that objective than policies optimized for another objective.\end{tabular}}} & \multicolumn{1}{l|}{\cellcolor[HTML]{EFEFEF}\textbf{learning}}                           & \multicolumn{1}{l|}{\cellcolor[HTML]{EFEFEF}\begin{tabular}[c]{@{}l@{}}\cmark* Supported for people low in NFC\\ \\ learning $>$ accuracy\end{tabular}}       & \cellcolor[HTML]{EFEFEF}\begin{tabular}[c]{@{}l@{}}\xmark \hspace{0.05cm} Not supported\\ \\ learning not significantly different from \\ accuracy\end{tabular}                                           \\ \hline
\end{tabular}

\caption{Summary of results for the main hypotheses tested in Experiment 1 \& 2. For a hypothesis: \cmark indicates support, \cmark * partial support, and \xmark \hspace{0.03cm} no support.}
\label{tbl:summary}
\end{table*}
\section{Experiment 1: Evaluating Participant Performance with Optimized Policies and SXAI}
\label{section:experiment-1}

The purpose of this study was to evaluate the effectiveness of the learned policies and an SXAI baseline in improving the respective objectives -- immediate decision accuracy and learning -- of people with different levels of NFC interacting with them. Table~\ref{tbl:summary} provides a combined summary of the findings from this experiment and from Experiment 2 (Section~\ref{section:experiment-2}).

\subsection{Hypotheses \& Research Questions}

Specifically, we hypothesized that:

\begin{itemize}

\item \textbf{H\hoptimalbaseline{b}:} Each NFC group interacting with RL policies optimized for specific target objectives will exhibit superior or comparable performance on those objectives when contrasted with individuals from the same NFC group interacting with the SXAI policy.

\item \textbf{H\hdiffobjective{a}:} Each group of NFC who interact with an RL policy optimized for immediate accuracy will perform better on immediate tasks compared to individuals from the same NFC group interacting with a policy that selects interactions based on the human learning.

\item \textbf{H\hdiffobjective{b}:} Each group of NFC who interact with an RL policy optimized for human learning will perform better on distal tasks (post-intervention questions) compared to individuals from the same NFC group interacting with a policy optimized for immediate decision accuracy.

\end{itemize}

In addition, we also sought to answer the following research questions with our work:

%\textbf{RQ1:} How will the policies personalized for different groups of NFC compare to a baseline explainable AI (XAI) approach on respective objectives?

\textbf{RQ3:} Will there be a trade-off between human learning (how much they learn) and their task enjoyment (including perceptions of effort required to perform the task)? A trade-off between effort and task enjoyment was previously observed in an AI-supported decision-making setting~\cite{bucinca2021trust}.

\textbf{RQ4:} Will greater learning of the task domain by the participants result in participants reporting higher perceived learning? Prior work indicates that perceived learning does not always reflect actual learning, particularly if environments that result in more learning require more effort~\cite{deslauriers2019measuring}.

\subsection{Task and conditions}

Participants were randomly assigned to one of the following conditions:
\begin{itemize}
    \item \textbf{Baseline, \em SXAI\em}. Participants were presented with AI recommendations and explanations in each question.
    \item \textbf{Accuracy}. Participants interacted with the policy optimized for immediate accuracy for their NFC group.
    \item \textbf{Learning}. Participants interacted with the policy optimized for learning for their NFC group.
\end{itemize}
Similar to the data collection study, participants completed a series of 33 questions, in which given a vignette about a fictitious character, they were tasked with selecting the optimal set of exercises about the character in question among two sets of exercises.
Each participant was again randomly assigned to three concepts. The total 33 questions were presented in three blocks (as shown in Fig~\ref{fig:experiment-flow}): \em pre\em, \em intervention\em,  \em post\em. \em Pre \em and \em post \em test blocks consisted of six questions each (two per concept), and the \em intervention \em block consisted of 21 questions (seven per concept).

\subsection{Experiment design}

The experiment design for the evaluation study consisted of three blocks: \em pre\em, \em intervention\em, and \em post\em. Participants received no AI assistance on the \em pre \em and \em post \em blocks, which served as test blocks to measure participants' learning. Participants interacted with one of the three policies (accuracy, learning, SXAI) during the intervention block.

The simulated AI system had an overall accuracy of ~71.4\%, with six out of 21 intervention questions having an underlying incorrect AI recommendation. As in the data collection study, the order of the questions was randomized for each participant and the questions in which the AI made incorrect suggestions were picked randomly. We ensured that AI had uniform accuracy across concepts: the six incorrect questions consisted of two incorrect questions per concept.

\subsection{Procedure}

The procedure was the same as for the data collection study. Participants were recruited via Prolific, a paid crowdsourcing platform and LabInTheWild.org. LabInTheWild~\cite{reinecke15:labinthewild} is an online platform where participants can voluntarily participate in a research study. Rather than receiving monetary compensation for participation, participants are presented with a detailed review of performance on the task and a comparison to other test-takers at the end of the study. %[[TODO]] %Previous work has validated...

\subsection{Participants}
347 participants were recruited and were randomized into the three reported policies (\em accuracy\em, \em learning\em, \em SXAI \em). 
We retained 316 participants who passed an attention check at the end of the study and demonstrated a median completion time of over four seconds per question, given that the questions involved reading a vignette. (Participants' demographics can be found in Appendix, Table~\ref{table:participants}). 
Each participant landing on the study from Prolific received compensation of 2.4 USD (12.72 USD/hr). Participants were assigned to policies optimized for their respective NFC groups. Categorization into \em low \em or \em high \em NFC groups depended on whether participants' scores fell within the lower or upper 50th percentile of the NFC scores obtained from the data collection study.

\subsection{Design and Analysis}

The study was a mixed between- and within-subjects design. There was one between-subjects factor, policy choice, with three levels: 1. the \em accuracy \em policy, 2. the \em learning \em policy, 3. \em SXAI\em.

The within-subjects factor was the concept, with participants interacting with three out of four possible concepts assigned to them randomly.

We collected the following objective measures:

\begin{itemize}
    \item \textbf{Immediate accuracy}: Percentage of correct answers in the \em intervention \em questions.
    \item \textbf{Learning}: Percentage of correct answers in the \em post \em questions (controlling for participant's performance in \em pre \em questions).
    \item \textbf{Overreliance}: Percentage of incorrect answers in questions where the AI was incorrect and participants received any type of AI assistance.

\end{itemize}

At the end of the study, we collected the following subjective measures, all on a 5-point Likert scale from 1=Strongly disagree to 5=Strongly agree:

\begin{itemize}
\item \textbf{Perceived learning}: Participants responded to \textit{``I believe I have learned about selecting exercises that are appropriate for a specific individual's goals, constraints, and preferences.''}
\item \textbf{Task enjoyment}: Participants responded to \textit{``I enjoyed this task.''}
\item \textbf{Mental demand}: Participants responded to \textit{``I found this task mentally demanding.''}
\item \textbf{Trust}: Participants responded to \textit{``I trust this AI's suggestions for optimal activities.''}

%%Intrinsic Motivation Inventory (IMI) https://selfdeterminationtheory.org/wp-content/uploads/2022/02/IMI_Complete.pdf

\end{itemize}

We used analysis of variance to analyze the impact of the different policies on both objective and subjective measures. The performance of participants on intervention questions was analyzed using mixed-effects models. The policy was modeled as a fixed effect, and the participant and concept in question as random effects. For analyzing learning, it is important to note that participants responded to only 6 post-intervention questions (2 for each concept). To ensure data conformity with a normal distribution, we employed analysis of variance on the average post-intervention question scores per participant, with policy as fixed effect and performance on pre-test questions as a covariate. Similarly, for the subjective measures, analysis of variance with policy as a fixed effect was used. We used Tukey's HSD for post-hoc comparisons.
For some of the hypotheses, our argument rests on the lack of differences between two policies. For such situations, we report $95 \%$ confidence intervals on the effect size (Cohen's $d$). If the reported interval spans 0 and the interval is narrow, this approach helps to show that, if any difference between the treatments (i.e., policies) exists, with high probability, it is small as the effect could be zero~\cite{colegrave2003confidence, lee2016alternatives, thompson2007effect}.

For the mixed-effect models, we report degrees of freedom obtained via Kenward-Roger method~\cite{kenward1997small}. 

\subsection{Results}

\begin{figure*}
     \centering
     \begin{subfigure}{0.49\textwidth}
         \centering
         \includegraphics[width=\textwidth]{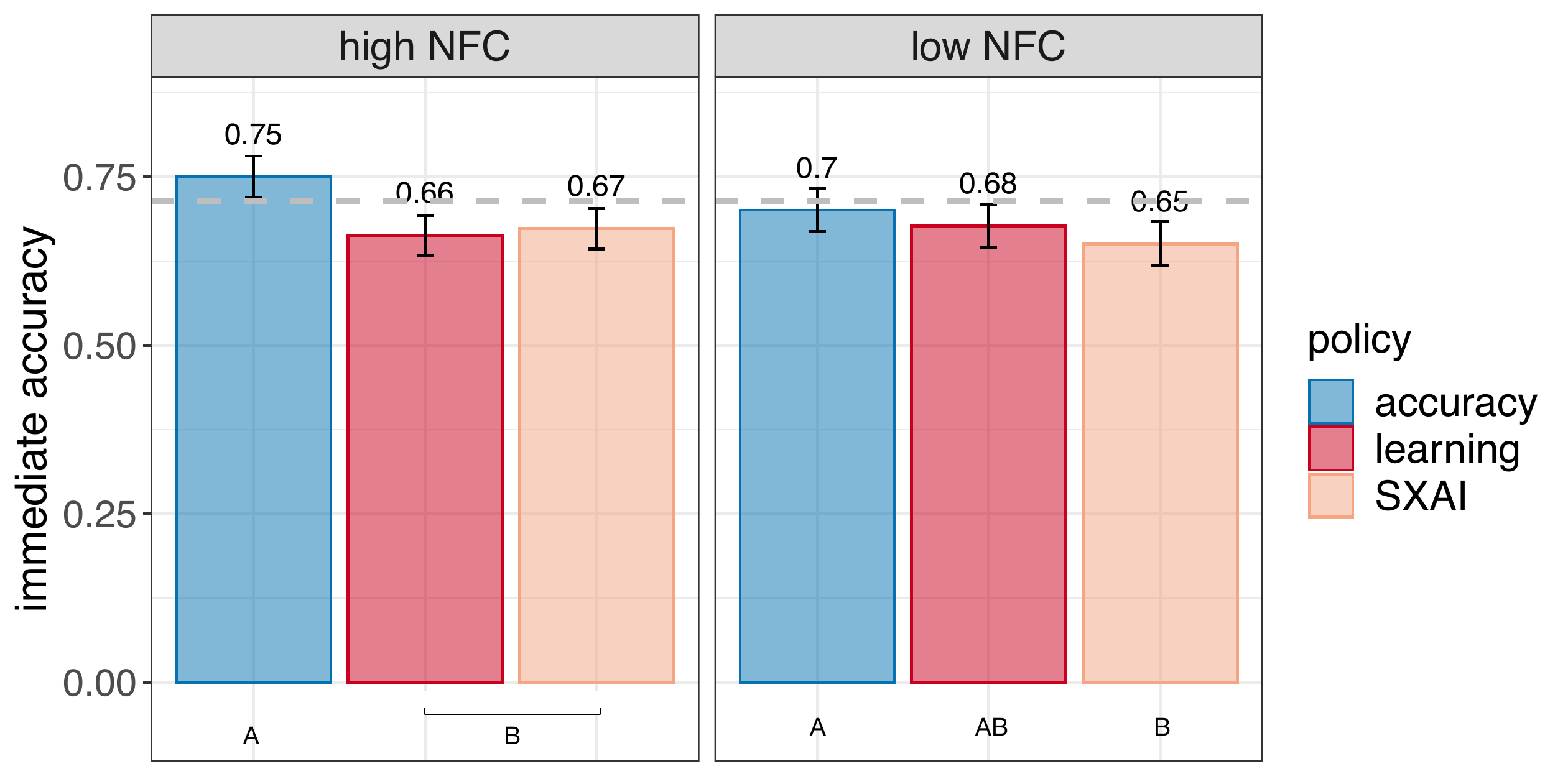}
         \caption{Immediate accuracy}
         \label{fig:pareto-accuracy-learning}
     \end{subfigure}
     \begin{subfigure}{0.49\textwidth}
         \centering
         \includegraphics[width=\textwidth]{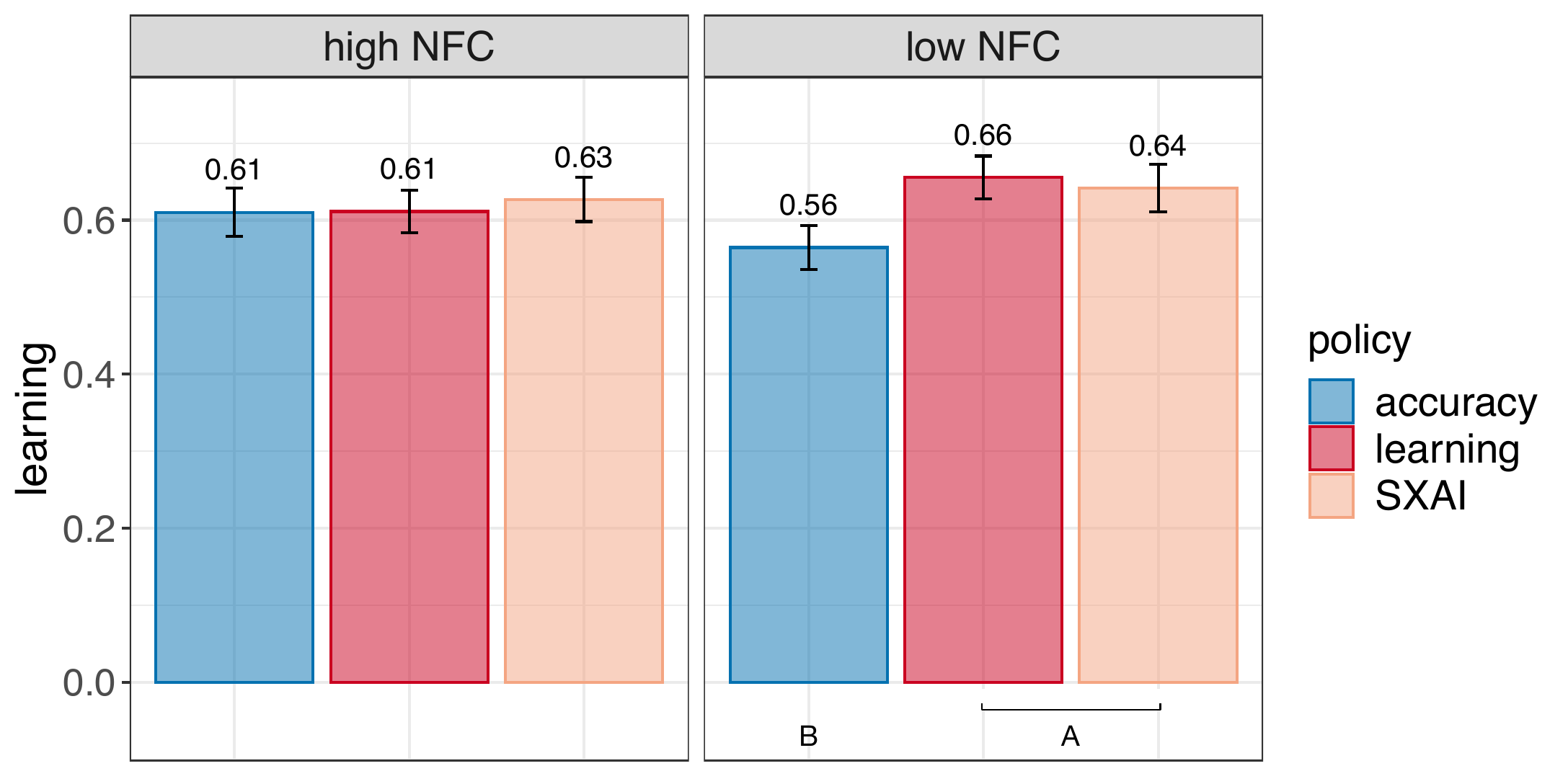}
         \caption{Learning}
         \label{fig:respective-accuracy-SXAI}
     \end{subfigure}

     \caption{Experiment 1: Marginal means of participants interacting with the three policies: \em accuracy\em, \em learning\em, \em SXAI\em,  on the two objectives: immediate accuracy and learning. Error bars indicate one standard error. The dashed line in (a) indicates the performance of the AI. Significance levels (if any) are depicted with letters. Conditions not connected by the same letter are significantly different.}
     \label{fig:main-result}
\end{figure*}

\subsubsection{Objective Measures}

Figure~\ref{fig:main-result} summarizes the main results. %\& Table~\ref{tbl:target-objectives} summarize the main results.

\textbf{Comparing optimized policies to the SXAI baseline on target objectives.} 
We investigated how the SXAI approach---in which participants were always presented with a recommendation and explanation---compared to the policies that were tailored to the NFC group and the objective of the interaction. Our results \textbf{support H\hoptimalbaseline{b}}. For accuracy as the objective, both NFC groups interacting with the policy optimized for immediate accuracy performed significantly better on immediate tasks than those interacting with SXAI (\em low NFC\em: $F_{1, 89.8} = 6.39, p = .01$, \em high NFC\em: $F_{1, 108.1} = 13.97, p = .0003$). For learning as the objective, both high and low NFC participants' performance on distal tasks with SXAI was not significantly better or worse than their performance on distal tasks when they interacted with the \em learning \em policy (\em low NFC\em: $F_{2, 92} = 0.06, p = .80 $, Cohen's $d = 0.06$, 95\% CI [-0.34, 0.48]; \em high NFC\em: $F_{2, 119} = 0.22, p = .64.$, Cohen's $d = - 0.07$, 95\% CI [-0.43, 0.28]).

\textbf{Comparing optimized policies to each other on target objectives.} 
First, focusing on \em accuracy \em as the objective, our results provide \textbf{partial support for H\hdiffobjective{a}}; it was supported for participants high in NFC, but we find no support for participants low in NFC. Participants high in NFC who interacted with the \em accuracy \em policy performed significantly better on immediate tasks than participants high in NFC who interacted with the \em learning \em policy ($F_{1, 111.2} = 16.73, p < .0001$). They also achieved complementary human-AI team accuracy ($M_{human} = 0.59, SE_{human} = 0.03$, $M_{human+AI} = 0.75, SE_{human+AI} = 0.03$; $ M_{AI}= 0.714; t(171) = 2.75, p = .006$). Participants low in NFC interacting with the \em accuracy \em policy did not perform significantly better (or worse) on immediate tasks compared to participants low in NFC who interacted with the \em learning \em policy ($F_{1, 97.65} = 1.59, n.s.$). 

Similarly, our results \textbf{partially support H\hdiffobjective{b}}, in which the focus is on the \em learning \em objective. In contrast to \textbf{H\hdiffobjective{a}}, \textbf{H\hdiffobjective{b}} was supported for participants low in NFC, but we find no support for participants high in NFC. Participants low in NFC who interacted with the \em learning \em policy performed significantly better on distal tasks (while controlling for their initial knowledge as measured during the pre-test) than participants low in NFC who interacted with the \em accuracy \em policy ($F_{2, 100} = 5.23, p = .02$). Whereas, participants high in NFC who interacted with the \em learning \em policy did not perform significantly better (or worse) on distal tasks compared to participants high in NFC who interacted with the \em accuracy \em policy ($F_{2, 87} = 0.10, p = n.s.$).

\begin{figure*}

\centering
\includegraphics[width=\textwidth]{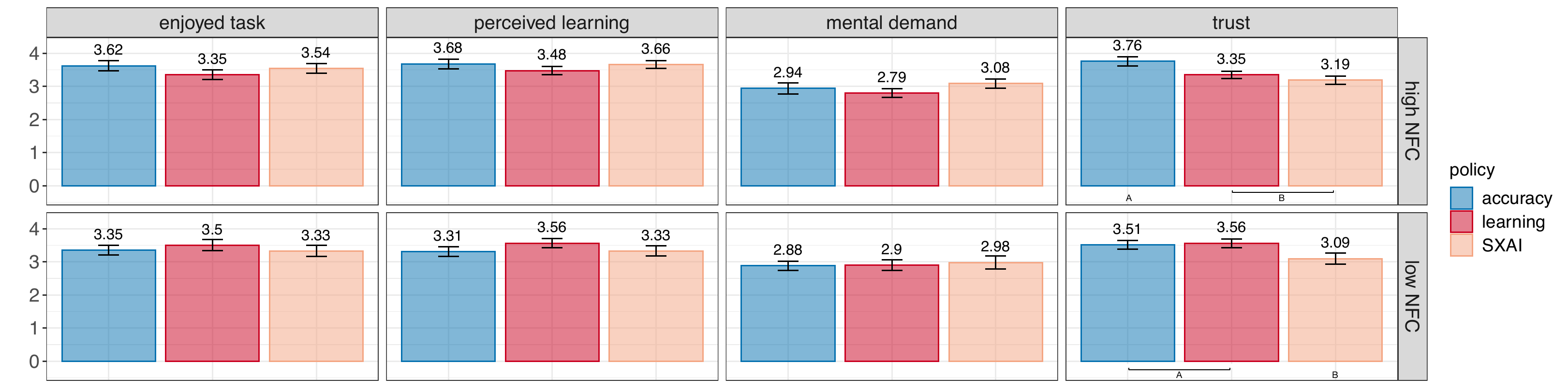}
\caption{Experiment 1: Subjective measures. Error bars indicate one standard error. Significance levels (if any) are depicted with letters. Conditions not sharing the same letter are significantly different.}
\label{fig:subjective-exp1}
\end{figure*}

\begin{figure*}

\centering
\includegraphics[width=\textwidth]{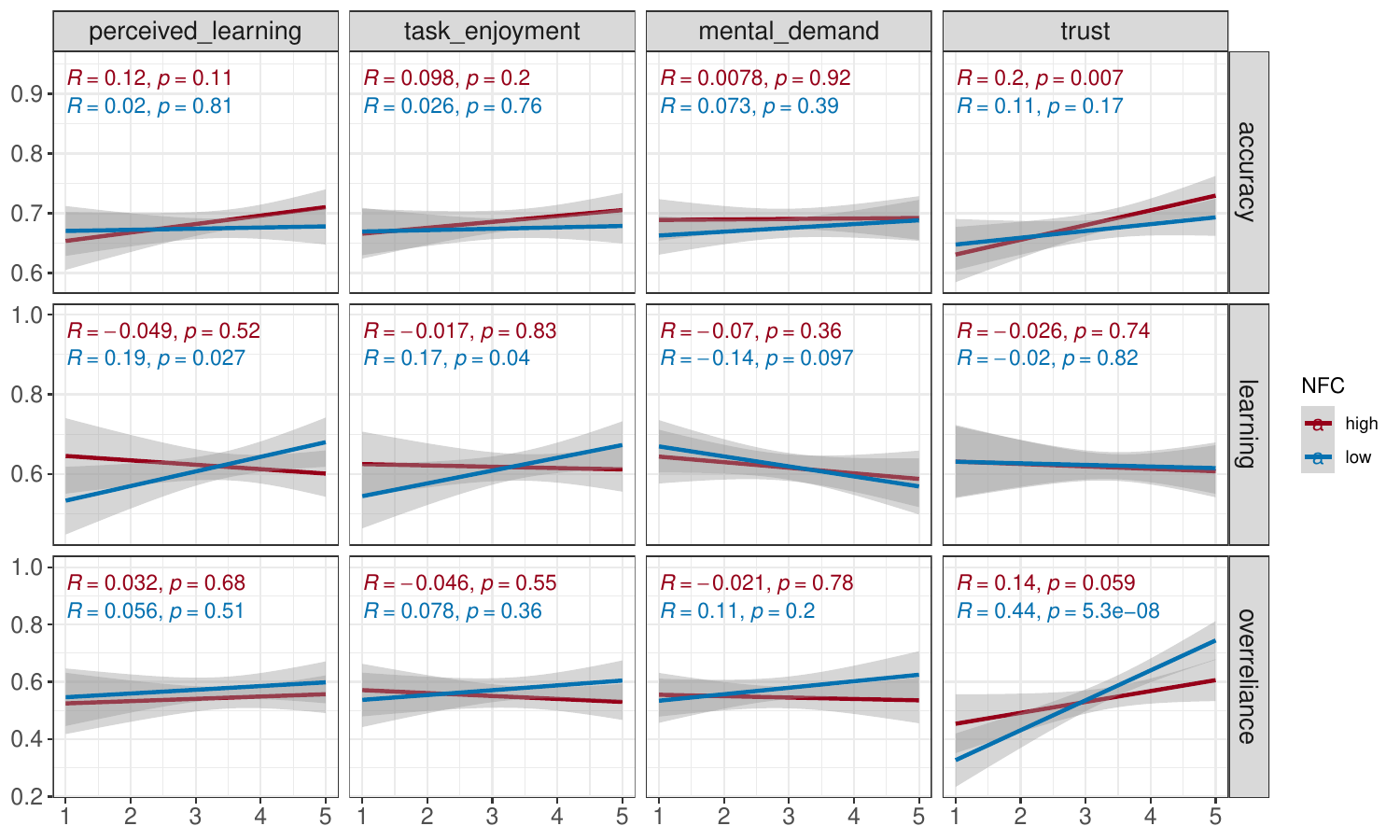}
\caption{Experiment 1: Relationships between objective vs. subjective measures for the two NFC groups.}
\label{fig:objective-vs-subjective}
\end{figure*}

\subsubsection{Subjective Measures}

We investigated the effect of policy on subjective ratings for each NFC group. Results for the subjective measures are summarized in Table~\ref{fig:subjective-exp1}. Each group trusted the policy that was more beneficial for them across objectives (i.e., the accuracy policy for people high in NFC, and the learning policy for people low in NFC) significantly more than the SXAI policy. Both groups also perceived that they had learned and enjoyed the task more with the policy that was more beneficial for them across objectives, but this trend was not significant.

\subsubsection{Objective measures vs. subjective measures}

Figure~\ref{fig:objective-vs-subjective} depicts the relationships between subjective measures and objective measures across policies for the two NFC groups.
Addressing \textbf{RQ3}, we do not observe a trade-off between actual learning and task enjoyment. In fact, task enjoyment was significantly positively correlated with actual learning for people low in NFC. We observed no substantial correlation for people high in NFC. There was also a significant positive correlation between actual learning and perceived learning for people low in NFC, answering \textbf{RQ4}. Trust was significantly correlated with immediate accuracy for people high in NFC. For both NFC groups, trust was positively correlated with overreliance, albeit for people high in NFC the correlation was marginal.

\subsection{Exploratory Analysis: Does Overreliance on AI Suggest a Lack of Cognitive Engagement?}

\label{section:exploratory-analysis}

In this section, we explore the relationship between overreliance and cognitive engagement, challenging the conventional assumption that overreliance results solely from a lack of cognitive engagement with the AI-provided information~\cite{bucinca2021trust, gajos2022people}.

Echoing past research, our computational analysis showed that for people low in NFC (i.e., those low in general cognitive motivation), the policy optimized to improve human learning included \em explanation only\em, an action that leads to cognitive engagement~\cite{gajos2022people}, as the top action (Figure ~\ref{fig:policy-distributions}). When evaluating participants' performance with the policies,  participants low in NFC who interacted with the learning policy indeed exhibited improved learning outcomes. This aligns with the notion that cognitive engagement is crucial for effective learning.

\begin{figure*}
     \centering
     
     \begin{subfigure}{0.3\textwidth}
         \centering
         \includegraphics[width=\textwidth]{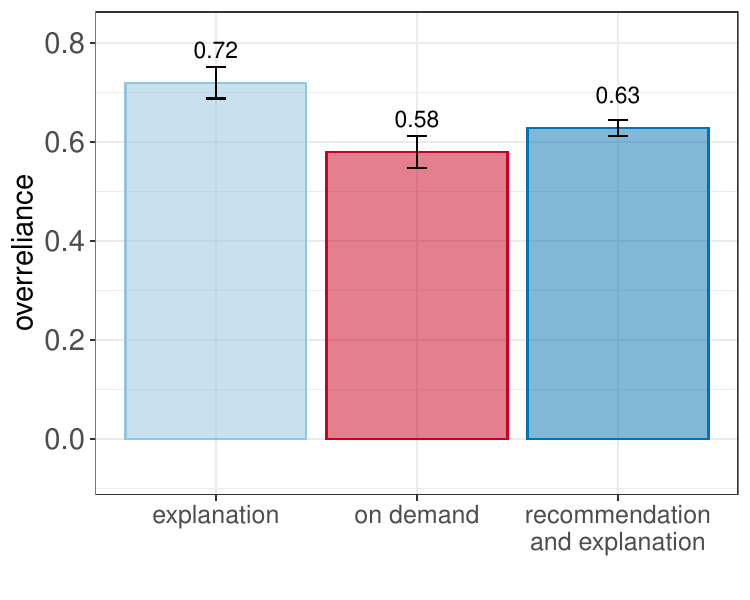}

         \includegraphics[width=\textwidth]{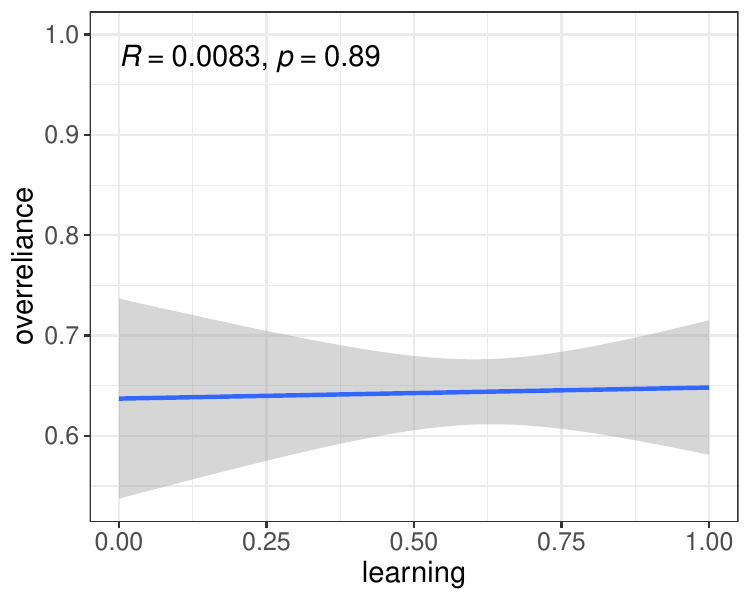}
         \caption{Experiment 1. The results of the exploratory analysis. Top: Overreliance by \em assistance type\em. Bottom: The relationship between overreliance and learning.}
         \label{fig:overreliance-by-assistance}
     \end{subfigure}
     \quad
     % \begin{subfigure}{0.3\textwidth}
     %     \centering
     %     \includegraphics[width=\textwidth]{figures/overreliance_vs_learning.pdf}
     %     \caption{}
     %     \label{fig:overreliance-vs-learning}
     % \end{subfigure}
     \begin{subfigure}{0.63\textwidth}
         \centering
         \includegraphics[width=\textwidth]{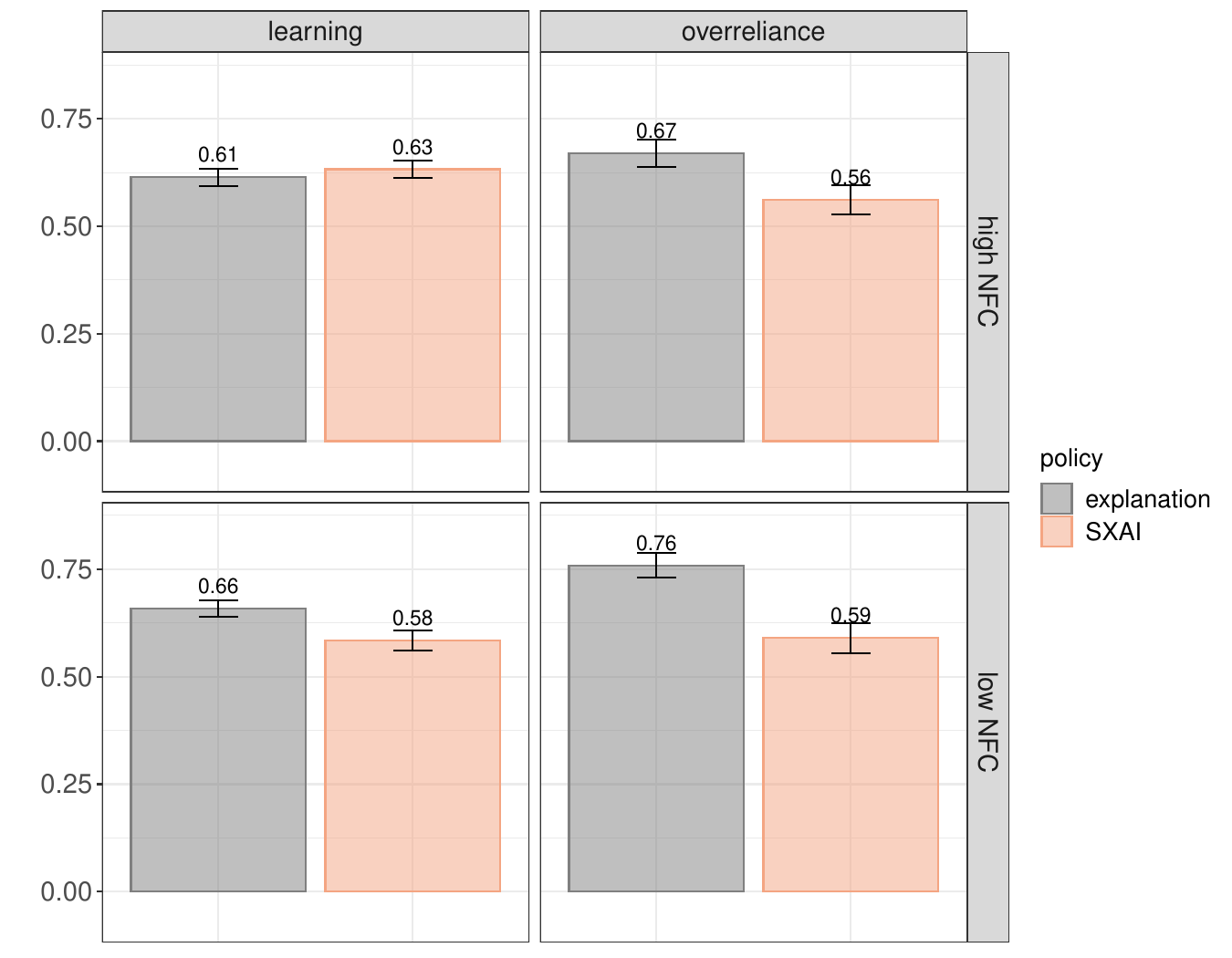}
         \caption{Experiment 2. Learning and overreliance by \em policy \em for both NFC groups. People low in NFC learned but also overrelied significantly more on the AI with the \em explanation \em policy compared to SXAI. (Error bars indicate one standard error.)}
         \label{fig:overreliance-vs-learning}
 
     \end{subfigure}

     \caption{Does Overreliance on AI Suggest a Lack of Cognitive Engagement?}
     \label{fig:exploratory-analysis}
\end{figure*}

However, a paradox emerged when we analyzed the effects of each assistance type individually (See Figure~\ref{fig:overreliance-by-assistance}). Surprisingly, \em explanation only \em assistance led to significantly more overreliance compared to other assistance types, including AI recommendation and explanation. Therefore, people both overrelied when making decisions with the \em explanation only \em action but also demonstrated enhanced learning when this action was predominant in the policy. This finding challenges the assumption that overreliance necessarily indicates a lack of cognitive engagement.

To further investigate the relationship between these two constructs, we conducted a correlation analysis between people's overreliance and their learning across policies (See Figure~\ref{fig:overreliance-vs-learning}). Given that incidental learning has been posited as a strong indicator of cognitive engagement, we would expect a negative correlation between learning and overreliance. However, we observed no substantial relationship between overreliance and learning $(r = 0.008, p = n.s., 95\% CI[-.11, .12])$. (Bootstrapped 95\% confidence intervals on $r$ were computed with 1000 bootstrapped datasets.) 
Based on this exploratory analysis, we included a new condition -- a fixed policy presenting explanations only -- in our subsequent experiment. We hypothesized that people will learn more but will also overrely more on AI when interacting with a fixed explanation only policy compared to another SXAI.
%This observation suggests that overreliance may result from factors beyond the depth of cognitive engagement and underscores the complexity of the phenomenon.

\section{Experiment 2: Evaluating Participant Performance with Optimized Policies and Multiple Baselines}
\label{section:experiment-2}

The second evaluation study had a primary and a secondary goal. Firstly, to further assess the effectiveness of our proposed approach ---  how policies optimized for different outcomes measure up against other baselines. Secondly, to test the new hypothesis about the relationship between overreliance and cognitive engagement that spurred from the exploratory analysis in Section~\ref{section:exploratory-analysis}.

To understand if it is possible to optimize participants' accuracy and learning jointly, we introduced a \em combined \em policy, an RL-based policy which was optimized for both accuracy and learning, by considering both immediate and distal rewards\footnote{We set $\gamma = 0$, equally weighing immediate and distal benefits with $\lambda$. Note that, in our setup, any $\gamma > 0$ would further weigh learning since future rewards might include accuracy attained with \em no AI \em support in the intervention phase.} ($\lambda = 0.5, \gamma = 0$). 
Alongside this, we included two additional baselines: \em explanation only\em, a fixed policy which provides only explanations in every decision and that prior work has shown to lead to learning without compromising accuracy~\cite{gajos2022people}, and a \em random policy \em which selects actions randomly on each question, as a mixed policy that does not account for contextual factors. 

\subsection{Hypotheses}

We adjusted \textbf{H\hoptimalbaseline{b}} to more broadly include the other baselines: 

\begin{itemize} 

    \item\textbf{H\hoptimalbaseline{b}:} Each NFC group interacting with RL policies optimized for specific target objectives will exhibit superior or comparable performance on those objectives when contrasted with individuals from the same NFC group interacting with a baseline policy.

\end{itemize}

Further, in addition to the hypotheses \textbf{H\hdiffobjective{a}} \& \textbf{H\hdiffobjective{b}}, we hypothesized the following for the combined policy:

\begin{itemize}
    \item\textbf{H\hdiffobjective{c}:} Each group of NFC who interact with RL policies optimized for both accuracy and learning (i.e., the \em combined \em policy) will perform similarly on the target objectives as the policies that were optimized \em solely \em for the target objective.
\end{itemize}

Informed by the exploratory analysis of cognitive engagement and overreliance, we additionally constructed the following hypothesis:

\begin{itemize}

    \item\textbf{H4:} Compared to the policy which provides AI recommendations and explanations in each decision (i.e., \em SXAI \em), the \em explanation only \em policy will lead to improved learning (as in~\cite{gajos2022people}), but also increased overreliance on AI.

\end{itemize}

\subsection{Task and conditions}

We used the same task design as in experiment 1. In this experiment, participants were randomly assigned to one of the following conditions:
\begin{itemize}
    \item \textbf{Baseline 1, \em SXAI\em}. Same as in Experiment 1 -- showing AI recommendation and explanation on each question.
    \item \textbf{Baseline 2, \em explanation only\em}.  In this condition, participants received \em explanation only \em as assistance for every question, a form of assistance previously demonstrated to enhance learning without compromising accuracy (only tested with correct AI recommendations)~\cite {gajos2022people}.
    \item \textbf{Baseline 3, \em random policy\em}. Participants were randomly provided one of four types of assistance for each question. We included this condition as a baseline to study the effect of variability of assistance on immediate accuracy and learning when that assistance is not selected by accounting for contextual factors.
    \item \textbf{Accuracy} Same as in Experiment 1 -- the policy optimized for immediate accuracy.
    \item \textbf{Learning} Same as in Experiment 1 -- the policy optimized for distal benefits.
    \item \textbf{Combined} In this condition, participants interacted with a policy that was optimized for both immediate accuracy and learning.
\end{itemize}

\subsection{Experiment design}

The experiment design was similar to Experiment 1 but modified to include more test questions and fewer intervention questions, aiming for a more reliable assessment of learning. The \em pre \em and \em post \em blocks consisted of 9 questions each, with 3 questions per concept. The \em intervention \em block consisted of 15 questions (5 per concept). The simulated AI system had an overall accuracy of ~73.33\%, with 4 out of 15 intervention questions having an underlying incorrect AI recommendation. These 4 incorrect questions consisted of the 3 concepts, with one concept being randomly chosen to be shown incorrectly twice per participant.

\subsection{Participants}
Out of 1063 recruited participants, 964 who passed the attention check and had a median completion time exceeding 4 seconds were retained for analysis. These participants were then randomly assigned to one of the six reported conditions (See Table~\ref{table:participants} in the Appendix for details). Each participant landing on the study from Prolific received compensation of 2.4 USD (10.02 USD/hour, with a median completion time of 14.29 minutes). As in Experiment 1, participants were assigned to policies optimized to their NFC levels (applicable for non-baseline conditions), categorized as \em low \em or \em high \em based on whether their scores were in the bottom or top 50th percentile from the data collection study.

\subsection{Procedure, Design and Analysis}
All the metrics and methods were the same as in Experiment 1.

\subsection{Results}

\begin{figure*}
     \centering
     \begin{subfigure}{0.49\textwidth}
         \centering
         \includegraphics[width=\textwidth]{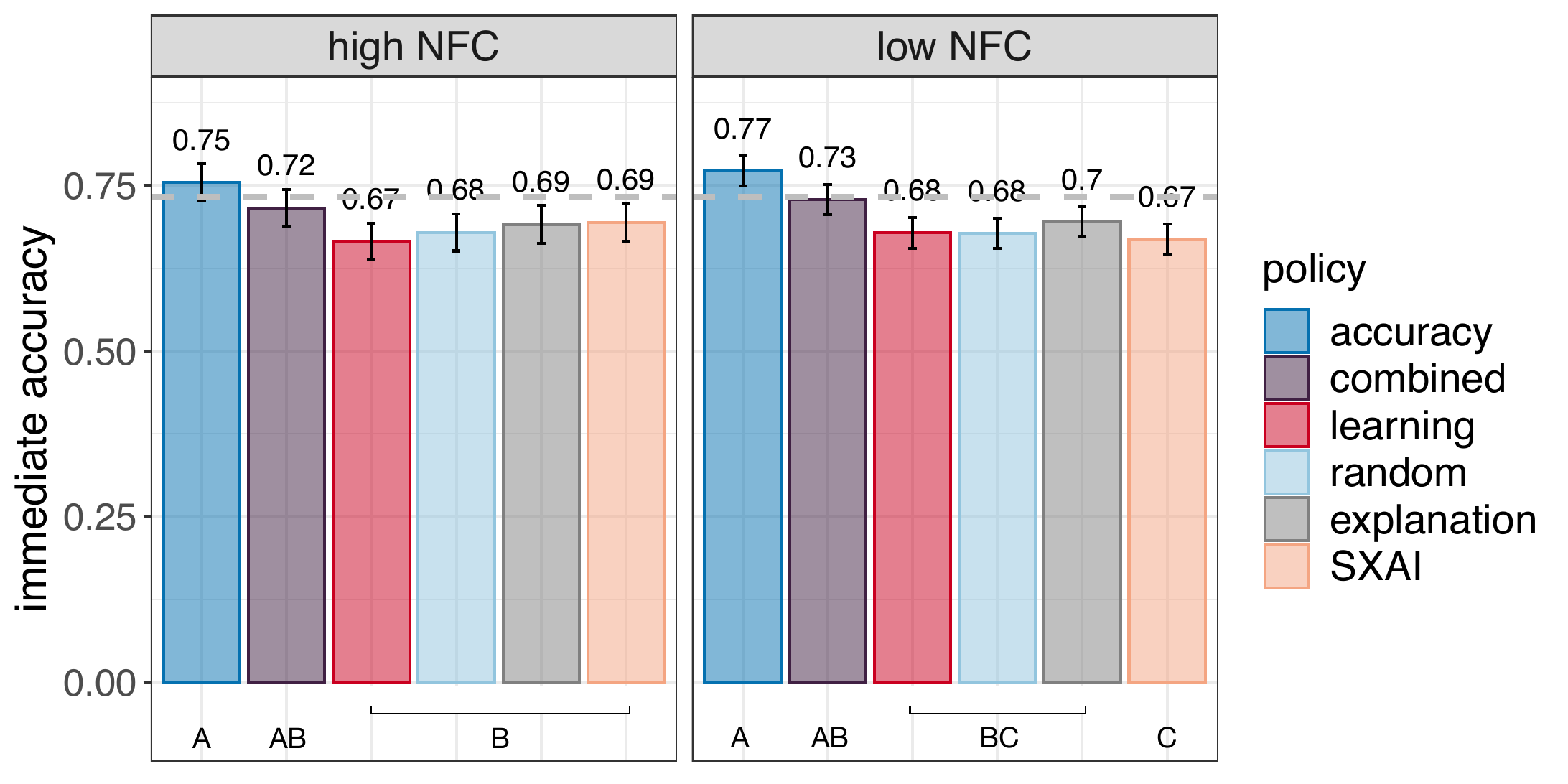}
         \caption{Immediate accuracy}
         \label{fig:accuracy-exp2}
     \end{subfigure}
     \begin{subfigure}{0.49\textwidth}
         \centering
         \includegraphics[width=\textwidth]{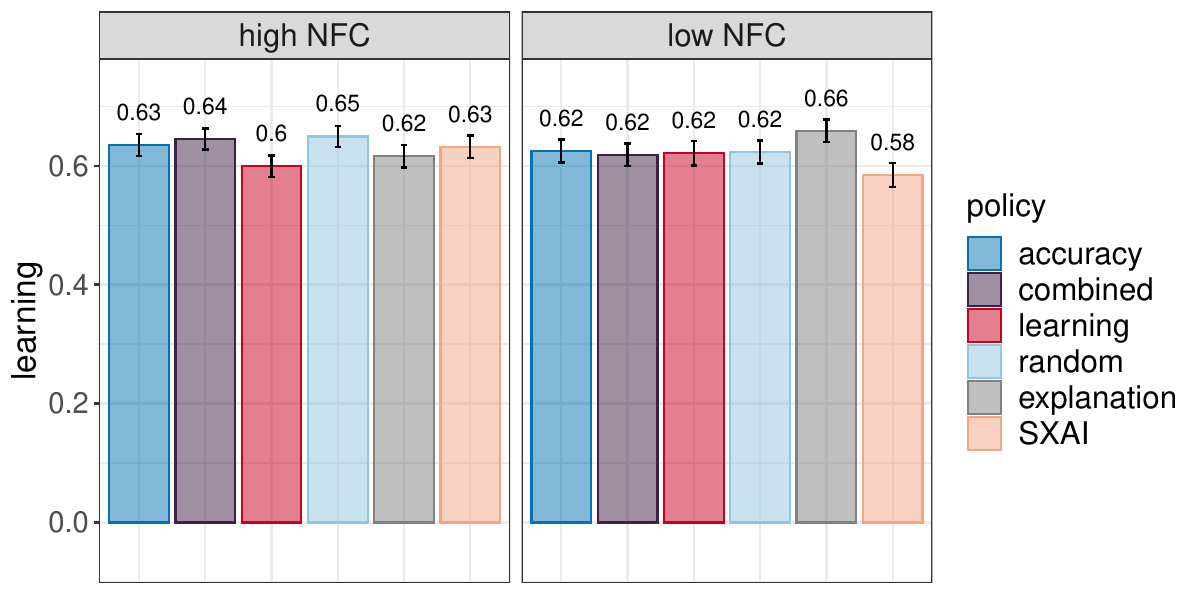}
         \caption{Learning}
         \label{fig:learning-exp2}
     \end{subfigure}

     \caption{Experiment 2: Performance of participants interacting with six policies on the two objectives: accuracy and learning. Error bars indicate one standard error. The dashed line in (a) indicates the performance of the AI. Significance levels (if any) are depicted with letters. Conditions not sharing the same letter are significantly different.}
     \label{fig:main-result-exp2}
\end{figure*}

\subsubsection{Objective Measures}

Figure~\ref{fig:main-result-exp2} summarizes the main results. The main effect of policy for \em immediate accuracy \em as the objective was significant for both groups of NFC (\em low NFC\em: $F_{5, 425.2} = 8.01, p < .0001.$, \em high NFC\em: $F_{5, 518} = 5.83, p < .0001$). Tukey's HSD comparisons are shown in Figure~\ref{fig:main-result-exp2}. Whereas, for learning as the objective the main effect was not significant for either group of NFC (\em low NFC\em: $F_{6, 434} = 1.23, n.s.$, \em high NFC\em: $F_{6, 528} = 1.48, n.s.$). 

\textbf{Comparing optimized policies to baselines on target objectives.} Our results \textbf{support H\hoptimalbaseline{b}}. For accuracy as the objective, both NFC groups interacting with the \em accuracy \em policy performed significantly better on immediate tasks than those interacting with any of the baselines: \em SXAI\em, \em explanation only\em, \em random \em policies. Participants low in NFC achieved human-AI complementary team performance ($M_{human+AI} = 0.77, SE_{human+AI} = 0.03$; $ M_{AI}= 0.73; t(89.67) = 2.52, p = .01$).
For learning as the objective, both high and low NFC participants' performance on distal tasks with baselines was not significantly better or worse than their performance on distal tasks with the \em learning \em or the \em combined \em policy (Pairwise comparisons with Tukey's HSD do not detect any significant differences among the policies. For pairwise effect sizes and confidence intervals see Appendix Table~\ref{table:effect_sizes}).

\textbf{Comparing optimized policies to each other on target objectives.}
Our results \textbf{support H\hdiffobjective{a}} for both NFC groups in this experiment. For each NFC group, participants who interacted with the policy that was optimized for immediate accuracy performed significantly better on immediate tasks than participants who interacted with the learning-optimized policy. 
In contrast to the first experiment, we \textbf{do not find support for H\hdiffobjective{b}}, with learning as the objective. There were no significant differences in performance on distal tasks between participants interacting with the \em learning \em policy and the participants interacting with the \em accuracy \em policy or \em combined \em policy, for either NFC group.

Lending \textbf{support to H\hdiffobjective{c}}, participants who interacted with the \em combined \em policy did not perform significantly better or worse (Tukey's HSD -- Figure~\ref{fig:main-result-exp2}) on either target objective compared to participants interacting with the policies that were optimized only for the accuracy objective (accuracy vs. combined --- \em high NFC\em: Cohen's $d=.08$, 95\% CI[-.02, .19], \em low NFC\em: Cohen's $d=.10$, 95\% CI[-.02, .22])  or only the learning objective (combined vs. learning --- \em high NFC\em: Cohen's $d=.27$, 95\% CI [-.02,.56]; \em low NFC\em: Cohen's $d=-.04$, 95\% CI [-.37,.29]).

%%%%% Subjective measures
\subsubsection{Subjective measures}
Results of subjective measures are summarized in Figure~\ref{fig:subjective-exp2}. For subjective measures, the main effect of policy was only significant for perceived learning for participants high in NFC. Participants high in NFC perceived to have learned significantly more with the \em accuracy \em and \em combined \em policy than with the \em SXAI\em, \em random \em or \em learning \em policies ($F_{5, 528} = 3.66, p = .003$).

\subsubsection{Objective measures vs. subjective measures}
As in Experiment 1, we do not observe a trade-off between actual learning and task enjoyment (\textbf{RQ3}). We further do not observe a correlation between actual and perceived learning, which was significant only for people low in NFC in Experiment 1 (\textbf{RQ4}). (To avoid repetition, the results of objective versus subjective measures from Experiment 2 are presented in the Appendix, Figure ~\ref{fig:subjective-exp2}.)

\begin{figure*}

\centering
\includegraphics[width=\textwidth]{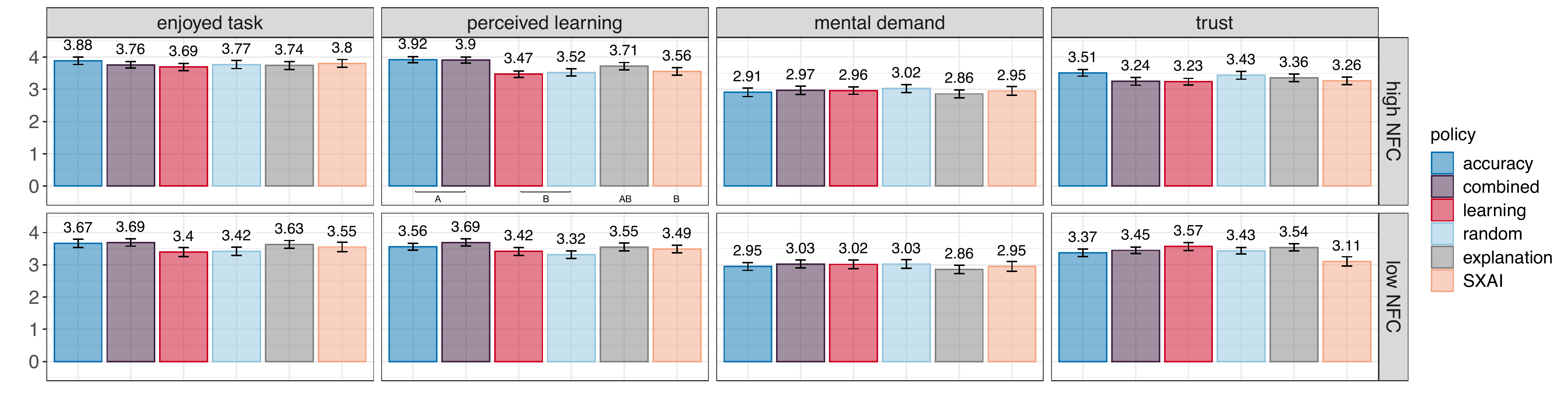}
\caption{Experiment 2: Subjective measures. Error bars indicate one standard error. Significance levels (if any) are depicted with letters. Conditions not sharing the same letter are significantly different.}
\label{fig:subjective-exp2}
\end{figure*}

%%% Cognitive engagement and overreliance

\subsubsection{Overreliance does not necessarily indicate a lack of cognitive engagement.}

Our results \textbf{supported H4} only for people low in NFC. 
Specifically, people low in NFC who interacted with the explanation policy learned significantly more than people low in NFC who interacted with the SXAI policy ($F_{2, 140} = 3.22, p = .04$). But they also overrelied significantly more on AI when interacting with the explanation policy compared to the SXAI policy ($F_{1, 163.3} = 4.48, p = .04$).

\section{Discussion}

In this work, we introduced offline reinforcement learning (RL) to learn decision support policies that optimize different human-centric objectives in AI-assisted decision-making. We instantiated our proposed approach with two objectives --- immediate accuracy of the decisions or long-term learning by human decision-makers --- while accounting for individual differences in people's need for cognition (or NFC, which reflects a person's intrinsic motivation to think) and other contextual factors. 

\subsection{Effectiveness of adaptive decision-making support for accuracy and learning as objectives}

Our work demonstrated that AI assistance needs to be dynamic, changing in response to context, individual differences, and the specified objective. In particular,  our results showed that better solutions than fixed policies like SXAI or mixed policies which do not consider these factors (i.e., random policy) can be learned for optimizing for immediate accuracy as the objective in human-AI decision-making. Both \em high \em and \em low \em NFC groups achieved significantly higher accuracy when they interacted with policies optimized for accuracy (i.e., \em accuracy \em  or \em combined \em policies) compared to interacting with (i) policies that do not adapt the support based on context, or (ii) policies that were optimized solely for the \em learning \em objective. Notably, participants high in NFC (in Experiment 1) and participants low in NFC (in Experiment 2) achieved complementary human-AI team performance with the accuracy policy, outperforming both human and AI accuracy alone on the task. 
Together with recent work~\cite{noti2022learning, ma2023should}, our results provide further evidence that adaptive interventions that consider contextual factors (e.g., AI's uncertainty, the decision-maker's competence or confidence) may be a promising approach for achieving the sought-after human-AI complementarity~\cite{bucinca20:proxy, bansal2021does}.

Our work also demonstrated that learning as an objective was more challenging to optimize than immediate decision accuracy. First, we observed that people high in NFC attained similar learning outcomes regardless of the policy they interacted with in both experiments, presumably because they were already motivated to engage with information. In contrast, those low in NFC benefited more from the learning policy than the accuracy policy in Experiment 1. However, Experiment 2 showed overall no significant difference in learning outcomes between these or other policies.  We believe the diminished learning signal across policies in Experiment 2, as opposed to Experiment 1, could be attributed to changes made in the experimental design. While the total number of questions remained constant across experiments, the second experiment had fewer intervention and more test questions. 
Although we aimed to strengthen learning measurement in Experiment 2 by increasing the number of test questions in our design update, we inadvertently limited exposure to concepts and, consequently, learning opportunities. 
We also believe that the task complexity, coupled with the explanation design, may have contributed to making learning challenging for the task. The task demanded not only an understanding of exercise-related facts, such as the benefits of pilates for flexibility, but also the skill to make suitable trade-offs, like balancing a fictitious character's goals against preferences. However, the explanations focused solely on exercise facts, and not of trade-offs, which may have added to the overall learning challenge. Our research underscores the critical need to develop robust explanations and human-AI interactions that consistently enhance human learning about the domain, as well as their accuracy on the task.

Comparing objective and subjective measures in AI-assisted decision-making, a trade-off between performance and preference was previously observed~\cite{bucinca2021trust}. Interestingly, our results showed that task enjoyment and perceived learning were positively correlated with actual learning for people low in NFC. No such correlation was observed for people high in NFC. A plausible explanation for this result may be that for low NFC participants, who are not generally inclined to engage in unnecessary cognitively-demanding activities, increased enjoyment of the task led to increased task-specific intrinsic motivation~\cite{deci2012self}, and that, in turn, led to greater cognitive engagement. 
These findings suggest that unlike previously assumed~\cite{bucinca2021trust}, cognitive engagement may not necessarily come at the cost of negative subjective experience.

\subsection{Evidence that RL is a promising approach for modeling human-AI decision-making }

Our computational analysis of the learned policies demonstrated that RL may be a valuable approach to modeling human-AI interaction in decision-making tasks. Specifically, we found that the composition of policies differed in meaningful ways depending on both the objective and the NFC group, and differed significantly from showing a fixed type of AI assistance (e.g., \em recommendation and explanation \em only). The analysis of the composition of these policies revealed some insights into how learning and immediate decision accuracy were supported for individuals with different levels of NFC.

\subsubsection{What have we learned about the impact of need for cognition?}

Our computational evaluation of the policies showed that the NFC group had a significant impact on shaping the composition of the policies optimized for learning, which necessitated cognitive engagement. 
Consistent with \textbf{H\hdiffnfc{}}, the learning policy favored actions that prior work has shown may promote cognitive engagement, such as \em explanation only\em~\cite{gajos2022people}, more often for individuals with low NFC than for those with high NFC. Considering that individuals with low NFC do not tend to naturally cognitively engage with information, which is critical for learning, this finding is in line with the NFC construct. 

NFC, however, was not a predictor of the policy composition for immediate decision accuracy (\textbf{RQ1}). An initial analysis indicated that the accuracy-optimized policies achieved their goal not by supporting cognitive engagement but by making reliance on the AI unlikely in those situations when the AI was incorrect. In those cases, AI assistance was often withheld. 
Specifically, when we examined the distributions of assistance types disaggregated by AI correctness (See Appendix, Figure~\ref{fig:accuracy-policy-by-AI-correctness}), we observed that when the AI was accurate (in half of the state space), the policies for both groups were nearly identical, predominantly favoring actions like \em recommendation and explanation\em, which prior research has indicated can lead to reliance, but also \em explanation \em only. Whereas, in cases where the AI was incorrect, the optimal course of action for individuals with high NFC clearly leaned toward \em no assistance\em, whereas for individuals with low NFC, the optimal choices included both \em no assistance \em and \em on demand\em. One possible explanation for why the policy might have discovered this signal is that individuals with low NFC tend to be less inclined to actively seek out information. Consequently, they were also less inclined to click and view the AI suggestion when it was presented in an \em on demand \em assistance format,  rendering  \em on demand \em  and  \em no assistance \em  similar interventions for them. To understand whether that was indeed what happened, we looked at how often individuals with low and high NFC clicked on the AI suggestion when it was offered in the \em on demand \em assistance format in the data collection study. Our findings revealed that individuals with low NFC clicked on the AI suggestion only 21\% of the time, whereas those with high NFC were notably more inclined to click on the suggestion, doing so 52\% of the time.
This finding expands our understanding of how different individuals interact with different assistance types and it also highlights the promise of computational models to enable discovery about human-AI decision-making.

\subsubsection{What have we learned about different objectives?}

Our inspection of the learned policies also revealed that the objective influenced the policy composition for both NFC groups. For people low in NFC, the \em learning \em policy favored actions that induced cognitive engagement (i.e., explanation), whereas the \em accuracy \em policy favored actions that induced \em reliance \em when AI was correct (i.e., recommendation and explanation). On the other hand, given that people high in NFC are already motivated to engage with information, \em recommendation and explanation \em was a good action for both \em accuracy \em and \em learning\em.  

When optimizing accuracy as the objective, \em no assistance \em was chosen more frequently than when the objective was learning. It was an optimal action for close to half of the state space, which corresponded to the instances when the AI was incorrect. This suggests that the policy recognized not providing assistance was the best option for preventing overreliance when the AI made mistakes, corroborating findings from previous research~\cite{noti2022learning}. In addition, given that \em no assistance \em was less often shown when optimizing learning, this also suggests that for improving learning any information is more beneficial than no information at all. Looking at the composition of the learning policy when the AI was correct vs. incorrect (See Figure~\ref{fig:learning-policy-by-AI-correctness}), we do not observe any differences in policy compositions, suggesting that AI correctness was irrelevant to supporting learning. We believe this result is due to our explanation design, which provided factually correct, albeit irrelevant, information when the AI was incorrect. Therefore, participants may have learned useful information even though the AI suggestion was wrong.

Interestingly, the distributions of actions for the \em combined \em policy, which was optimized for both learning and accuracy were similar to the learning policy for people low in NFC, and similar to the accuracy policy for people high in NFC (See Appendix, Figure ~\ref{fig:combined-policy}). For both groups, the respective policies were the better choices across objectives (in Experiment 1). This finding further suggests that the policies are optimizing for both objectives successfully, albeit the learning signal is more difficult to capture.

\subsection{Does overreliance on AI suggest a lack of cognitive engagement?}

Appropriate reliance on AI and cognitive engagement are critical constructs to consider when seeking to optimize accuracy and learning in AI-assisted decision-making. Fractured evidence from prior work suggested that there exists a relationship between cognitive engagement and overreliance on AI. Some assistance types such as \em recommendation and explanation \em induced reliance on AI, regardless of AI correctness~\cite{bucinca2021trust}.  But they also led to no learning about the domain~\cite{gajos2022people}. %Thus, such assistance types could be useful in improving human-AI accuracy when the AI had high confidence. 
Other assistance types, such as providing an explanation only without a decision recommendation, improved learning about the domain (indicating cognitive engagement)~\cite{gajos2022people}, but their effect on (over)reliance was not evaluated.  The underlying explanation for the difference in learning outcomes (and the possible difference in reliance) for the two assistance types was related to the effect they had on cognitive engagement. It was suggested that cognitive engagement explained both learning and overreliance. Assistance types that induce cognitive engagement lead to learning \em and \em should reduce overreliance on AI~\cite{gajos2022people}.

Our results paint a more complex picture. Drawing on exploratory analysis (\S~\ref{section:exploratory-analysis}) and the subsequent finding (\S~\ref{section:experiment-2}) that \em explanation only \em improved learning compared to \em SXAI \em, we believe that cognitive engagement does indeed improve learning. But contrary to the field's tentative understanding, lack of cognitive engagement may not be the sole predictor of overreliance.  The first piece of evidence that supports this hypothesis is that in our exploratory analysis, we find no substantial correlation between overreliance and learning across policies. Also, when analyzing the effect of individual assistance types on overreliance, we observed that \em explanation only \em led to significant overreliance compared to other assistance types, but people interacting with policies where \em explanation only \em was the predominant assistance type also exhibited improved learning. Decisively, Experiment 2 revealed that people low in NFC both learned more \em and \em overrelied more on AI when interacting with \em explanation only \em policy compared to SXAI. 

Together our findings demonstrate that the relationship between overreliance and cognitive engagement is multifaceted and not as straightforward as previously assumed. While cognitive engagement remains a crucial aspect of learning, overreliance on AI may stem from various factors. It may be a result of superficial engagement with AI-provided assistance, but it may also be due to people engaging with the provided assistance \em and \em finding the AI's (misleading) explanation plausible.
Understanding these nuances is essential for developing effective AI systems that enhance learning and promote appropriate reliance.

\subsection{Generalizability \& Limitations}

%generalizability

We believe offline RL has the potential to optimize various human-centric objectives effectively in AI decision-making, although we focused only on accuracy and learning in this work. Success depends, however, on intelligently designing RL components (state space, action space, rewards) based on each specific objective. This can be especially challenging for human-centric objectives other than accuracy for which empirical evidence about effective assistance types and relevant factors is still scarce in AI-assisted decision-making literature, as we observed for learning as an objective.
Although we used offline RL in this work (we ran an initial study with an exploratory policy, and then optimized for policies), we could have run an online RL algorithm instead (where we learn optimal policies during a study). However, this can be risky (by taking exploratory unsafe actions in real-time in the environment) and computationally expensive (especially when decisions are required in time-constrained settings). We used Q-learning as our offline off-policy learning algorithm: other algorithms are possible, but Q-learning was sufficient in our relatively simple (discrete) state and action-space. 

%Q-learning as an offline model-free algorithm that does not necessitate explicitly building a model of the environment's dynamics.
%add how this compares to other RL algorithms

Our work has several limitations.
Our findings are based on studies conducted with a single task, in a non-critical domain, and with crowds. However, since previous research in AI-assisted decision-making has demonstrated that experts exhibit behavior akin to that of crowds when utilizing AI for decision-making~\cite{gaube2021ai}, we may expect our findings to generalize in real contexts with experts making decisions in critical domains.
Our results may have also partially been driven by the explanation choice, which was explicitly designed to enable learning of factual information about the task domain (e.g., swimming supports muscle building). As is the case with accuracy~\cite{chen2023understanding, bansal2021does, bucinca20:proxy}, we believe that different explanation designs may have different impacts on learning.

\section{Ethical Considerations}

The status quo of deploying AI decision support systems without understanding their impact on people --- their skills, enjoyment, autonomy, collaboration with others, and work meaning --- is implicitly a value-laden decision. While our approach introduces a novel research direction focused on making such values explicit by enabling optimization of human-centric objectives in AI-assisted decision-making, it also surfaces ethical issues that must be addressed. One critical aspect is deploying technologies in a worker-centered way and ensuring that individuals engaging with these systems possess the autonomy to shape the influence these technologies have on them and their work environment~\cite{kawakami2023sensing, adler2022burnout, ajunwa2020black}. Therefore, it is critical for the system objectives to be determined and inputted by the users themselves, rather than being paternalistically imposed by those in power 
(e.g., managers). Further, personalization variables in the algorithm based on factors like skill level or motivation to think, may be used for nefarious purposes and could lead to unfair treatment or discrimination in workplace settings. It is essential to safeguard user privacy of such variables in the system design and allow only individuals interacting with the system to decide what variables can be tracked and utilized for personalization~\cite{asthana2024know, ajunwa2020black}.
%personalization for nefarious purposes

\section{Conclusion}

We proposed offline RL as an approach to provide adaptive support for optimizing different human-centric objectives in AI-assisted decision-making. To instantiate our approach, we considered decision accuracy and learning about the domain as objectives that are important to optimize. We constructed the state space and action space according to these objectives and learned decision-support policies. Our results showed that our approach was consistently successful in improving the accuracy of people interacting with the optimized policies. Whereas learning was more difficult to optimize, leading to improved outcomes only in specific instances.
Overall, our results demonstrated that offline RL is a promising approach for modeling human-AI decision-making, leading to both policies that optimize objectives and through interpretation, can reveal novel insights about human-AI decision-making space. Our research also underscores the importance of considering human-centric objectives beyond decision accuracy in AI-assisted decision-making and identifies the open research area for designing human-AI interaction that improves learning and other human-centric objectives along with accuracy.

\section*{Acknowledgements}
This work was supported in part by the National Science Foundation under Grant No. IIS-2107391. Any opinions, findings, and conclusions or recommendations expressed in this material are those of the author(s) and do not necessarily reflect the views of the National Science Foundation. We thank Daniel Oppenheimer, Ian Arawjo, Markus Langer, Eura Shin, Katy Gero, Suzanne Smith, Yaniv Yacoby, Alex Cabral,  Harvard HCI and DtAK Labs, and the d3center at UMich for helpful suggestions and discussions. ZB was partially supported by an IBM PhD Fellowship.

\balance{}

% BALANCE COLUMNS
\balance{}

% REFERENCES FORMAT
% References must be the same font size as other body text.
\bibliographystyle{SIGCHI-Reference-Format}
\bibliography{sample,kzg}
\newpage
\appendix
\onecolumn
\section{Appendix}

\subsection{Participants' demographics}

\begin{table*}[h]
\begin{tabular}{l|lll}
\toprule
\textbf{}       & \textbf{Data Collection Study}& \textbf{Experiment 1}& \textbf{Experiment 2}\\
\toprule
\textbf{n}      & 142& 316                                                                                                                   & 964    \\
\midrule
\textbf{\begin{tabular}[c]{@{}l@{}}Data collection\end{tabular}}      & June 2023& July-August 2023                                                                                                                   & November 2023    \\
\midrule
\textbf{Source} & \begin{tabular}[c]{@{}l@{}}Prolific: 142\end{tabular}                                                   & \begin{tabular}[c]{@{}l@{}}Prolific: 281\\ LabintheWild: 35\end{tabular}                                                  & \begin{tabular}[c]{@{}l@{}}Prolific: 952\\ LabintheWild: 12\end{tabular} 
\\
\midrule
\textbf{Age}    & M=38.09, SD=13.59& M=38.84, SD=14.9                                                                                                                     & M=42.33, SD=14.47     \\
\midrule
\textbf{Gender} & \begin{tabular}[c]{@{}l@{}}women: 79\\ men: 63\end{tabular}& \begin{tabular}[c]{@{}l@{}}women: 170\\ men: 135\\ non-binary: 7\\ not responded: 4\end{tabular} & \begin{tabular}[c]{@{}l@{}}women: 493\\ men: 438\\ non-binary: 26\\ not responded: 7\end{tabular} \\
\midrule
\textbf{\begin{tabular}[c]{@{}l@{}}Conditions\\ (high NFC, low NFC)\end{tabular}}    &  \begin{tabular}[c]{@{}l@{}}exploratory policy: 142\\ (high: 74, low: 68)\end{tabular}& \begin{tabular}[c]{@{}l@{}}SXAI: 102 (high: 59, low: 43)\\ accuracy: 99 (high: 50, low: 49)\\ learning: 115 (high: 63, low: 52)\end{tabular} & \begin{tabular}[c]{@{}l@{}}SXAI: 146 (high: 81, low: 65)\\ accuracy: 161 (high: 86, low: 75)\\ learning: 159 (high: 94, low: 65)\\ combined: 172 (high: 94, low: 78)\\ explanation: 160 (high: 84, low: 76)\\ random: 166 (high: 90, low: 76)\end{tabular} \\
\bottomrule
\end{tabular}
\caption{Participants' demographics and assignment to conditions in the respective study.}
\label{table:participants}
\end{table*}

\subsection{Experiment 2: Relationships between subjective and objective measures}
\begin{figure}[h]
    \centering
    \includegraphics[width=0.7\textwidth]{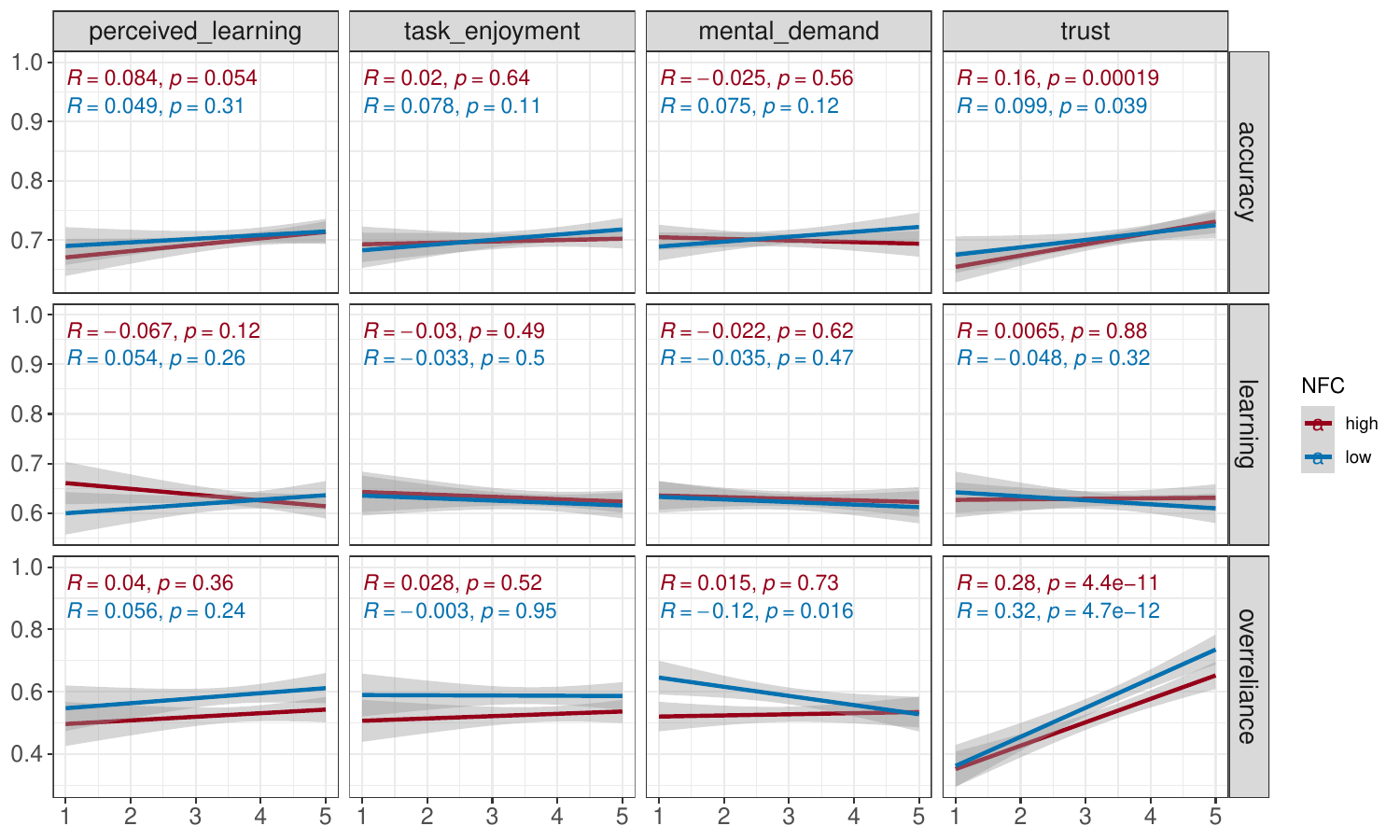}
    \caption{Experiment 2: Relationships between subjective measures and objective measures}
    \label{fig:objective-vs-subjective-exp2}
\end{figure}

\subsection{Distributions of actions for the optimized policies}

\begin{figure*}[hb]
    \centering
    \includegraphics[width=0.7\textwidth]{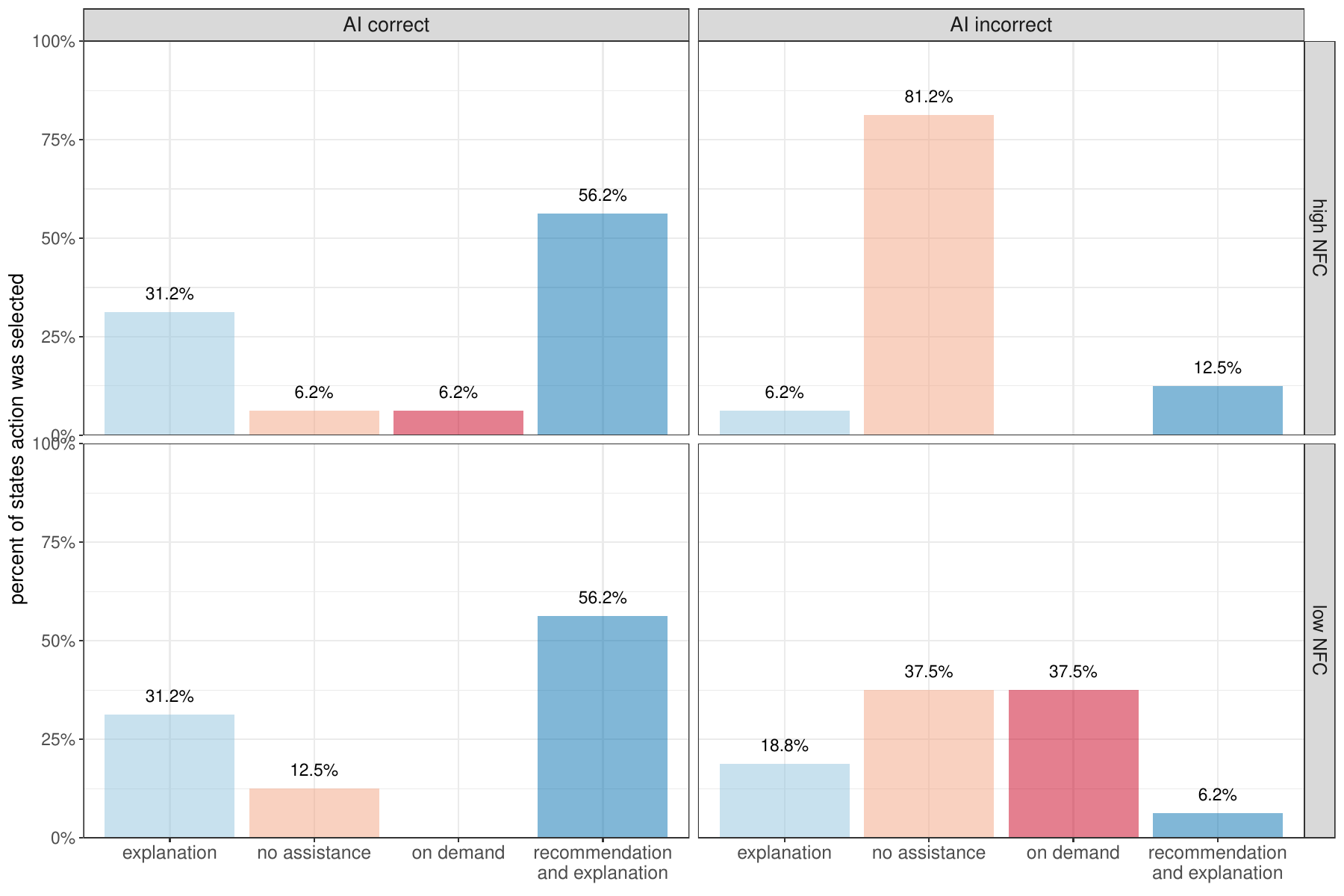}
    \caption{Distributions of types of AI assistance selected by the \em accuracy \em policy when the AI was correct and incorrect for different NFC groups.}
    \label{fig:accuracy-policy-by-AI-correctness}
\end{figure*}

\begin{figure*}[hb]
    \centering
    \includegraphics[width=0.7\textwidth]{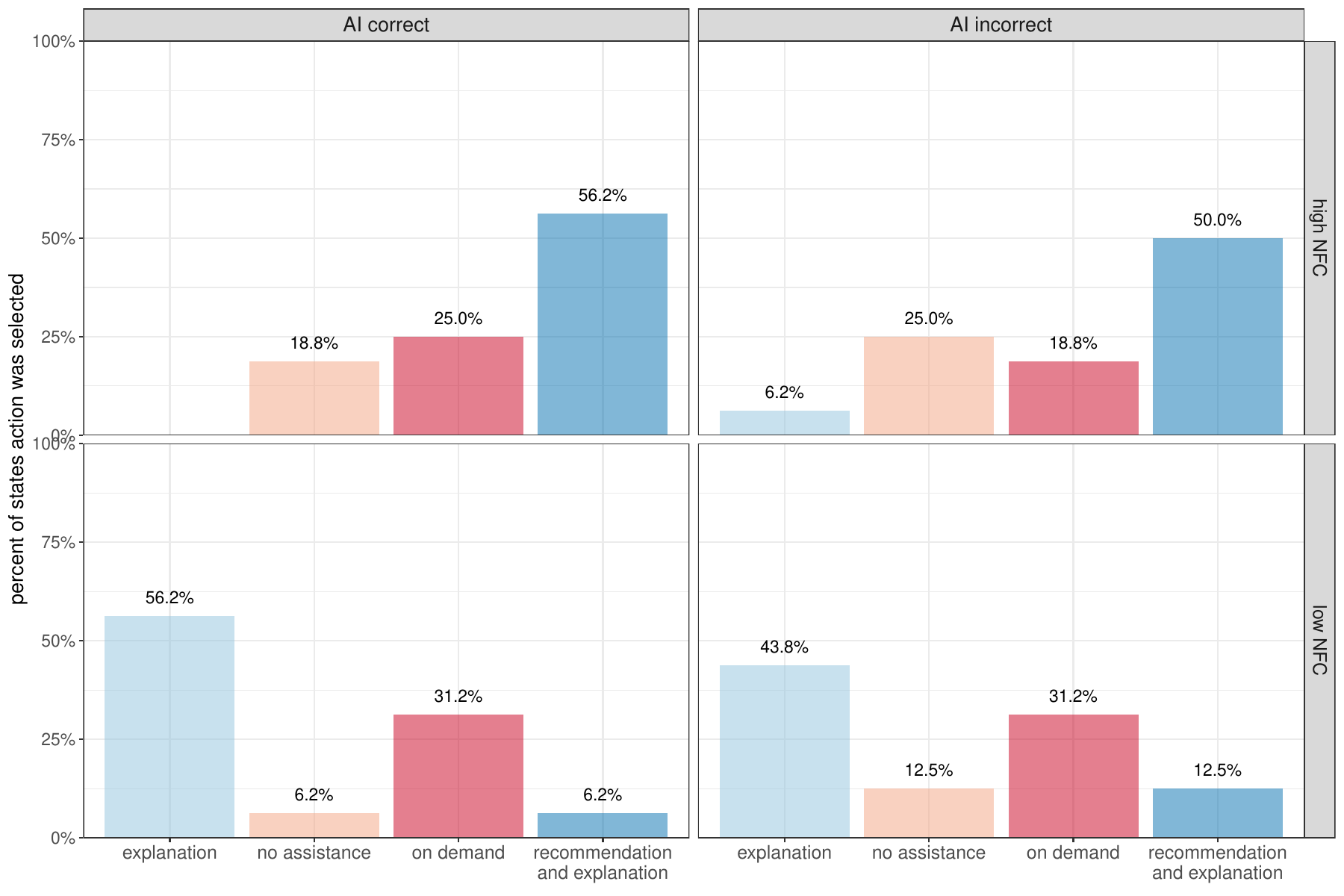}
    \caption{Distributions of types of AI assistance selected by the \em learning \em policy when the AI was correct and incorrect for different NFC groups.}
    \label{fig:learning-policy-by-AI-correctness}
\end{figure*}

\begin{figure*}[ht]
    \centering
    \includegraphics[width=\textwidth]{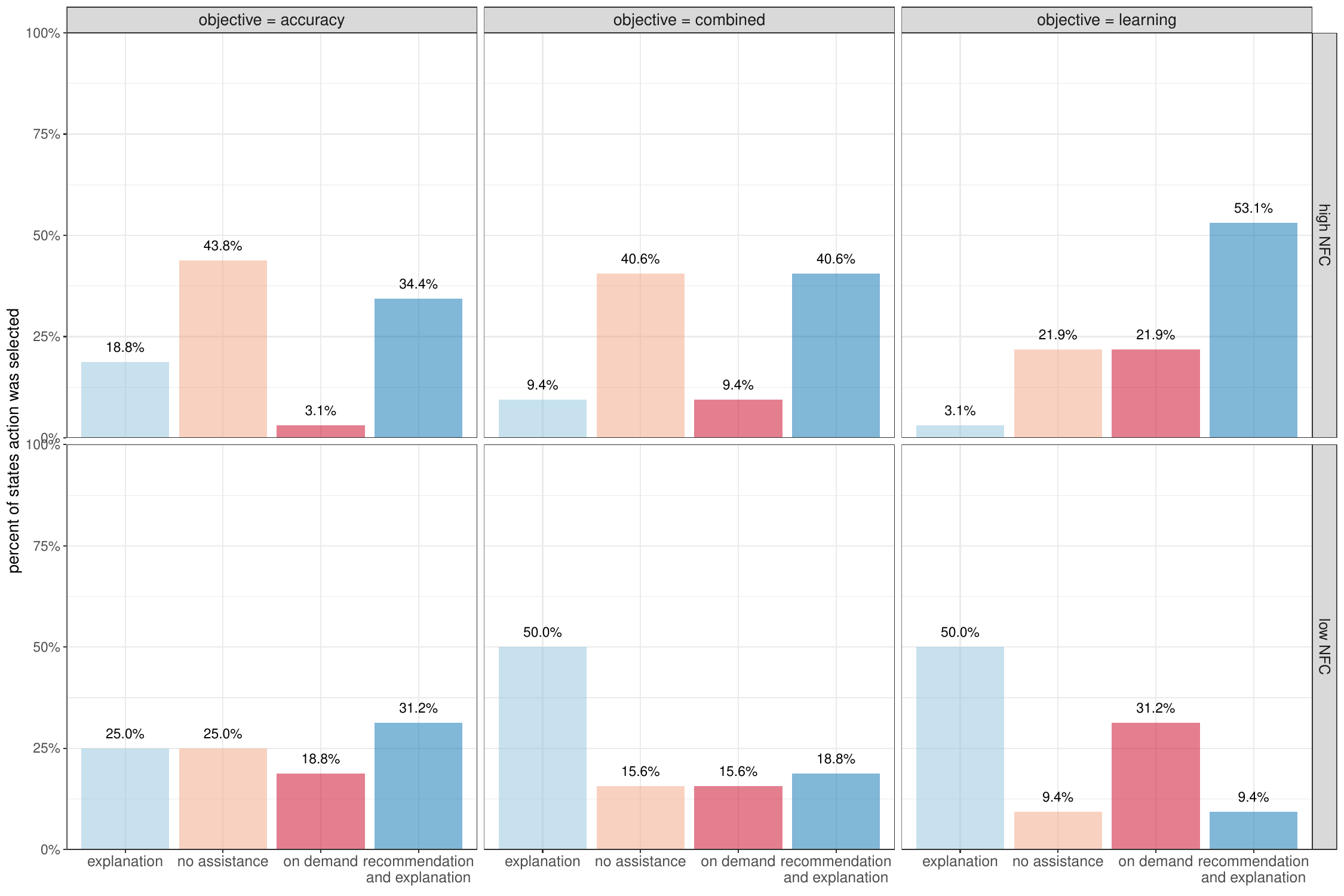}
    \caption{Distributions of types of AI assistance for the three policies \em accuracy \em, \em combined (accuracy + learning) \em and \em learning \em for different NFC groups.}
    \label{fig:combined-policy}
\end{figure*}
\newpage

\subsection{Experiment 2: Effect sizes for non-significant differences}
\begin{table}[H]
\begin{tabular}{@{}lll@{}}
\toprule
                                          & \textbf{high NFC}                                                                         & \textbf{low NFC}                                                     \\ \midrule
\multicolumn{1}{l|}{\textbf{comparison}}  & \multicolumn{1}{l|}{\textbf{\begin{tabular}[c]{@{}l@{}}Cohen's d\\ 95\% CI\end{tabular}}} & \textbf{\begin{tabular}[c]{@{}l@{}}Cohen's d\\ 95\% CI\end{tabular}} \\ \midrule
\multicolumn{1}{l|}{learning-SXAI}        & \multicolumn{1}{l|}{\begin{tabular}[c]{@{}l@{}}-.18\\ {[}-.48, .11{]}\end{tabular}}       & \begin{tabular}[c]{@{}l@{}}.19\\ {[}-.15, .54{]}\end{tabular}        \\
\multicolumn{1}{l|}{combined-SXAI}        & \multicolumn{1}{l|}{\begin{tabular}[c]{@{}l@{}}.08\\ {[}-.22, .38{]}\end{tabular}}        & \begin{tabular}[c]{@{}l@{}}.17\\ {[}-.16, .51{]}\end{tabular}        \\
\multicolumn{1}{l|}{learning-random}      & \multicolumn{1}{l|}{\begin{tabular}[c]{@{}l@{}}-.29\\ {[}-.58, .006{]}\end{tabular}}      & \begin{tabular}[c]{@{}l@{}}-.02\\ {[}-.36, .31{]}\end{tabular}       \\
\multicolumn{1}{l|}{combined-random}      & \multicolumn{1}{l|}{\begin{tabular}[c]{@{}l@{}}-.03\\ {[}-.32, .26{]}\end{tabular}}       & \begin{tabular}[c]{@{}l@{}}-.05\\ {[}-.37, .27{]}\end{tabular}       \\
\multicolumn{1}{l|}{explanation-learning} & \multicolumn{1}{l|}{\begin{tabular}[c]{@{}l@{}}.07\\ {[}-.22, .37{]}\end{tabular}}        & \begin{tabular}[c]{@{}l@{}}.22\\ {[}-.12, .55{]}\end{tabular}        \\
\multicolumn{1}{l|}{combined-explanation} & \multicolumn{1}{l|}{\begin{tabular}[c]{@{}l@{}}.17\\ {[}-.13, .47{]}\end{tabular}}        & \begin{tabular}[c]{@{}l@{}}-.24\\ {[}-.56, .08{]}\end{tabular}       \\ \bottomrule
\end{tabular}
\caption{Effect sizes for H1b in Experiment 2.}
\label{table:effect_sizes}
\end{table}

\end{document}